\documentstyle[twocolumn,epsfig]{mn}
\def\spose#1{\hbox to 0pt{#1\hss}}
\def\simlt{\mathrel{\spose{\lower 3pt\hbox{$\mathchar"218$}}
     \raise 2.0pt\hbox{$\mathchar"13C$}}}
\def\simgt{\mathrel{\spose{\lower 3pt\hbox{$\mathchar"218$}}
     \raise 2.0pt\hbox{$\mathchar"13E$}}}

\def\pc{{\rm pc}}
\def\kpc{{\rm kpc}}
\def\lw{L_{\rm w}}
\def\lb{L_{\rm b}}
\def\ln{L_{\rm n}}

\def\nn{{\cal\char'116}}

\title{Dynamical and chemical evolution of gas-rich dwarf galaxies}
\author[Recchi et al.]
{Simone Recchi$^{1,2}$\thanks{E-mail: recchi@sissa.it (SR); 
francesc@sissa.it (FM); annibale@astbo3.bo.astro.it (AD)}, 
Francesca Matteucci$^{1,2}$ and 
Annibale D'Ercole$^{3}$ \\
$^1$Dipartimento di Astronomia, Universit\`a di Trieste, Via G.B. Tiepolo, 11,
34131 Trieste, Italy \\
$^2$SISSA/ISAS, Via Beirut 2-4, 34014 Trieste, Italy \\
$^3$Osservatorio Astronomico di Bologna, via Ranzani 1, 44127 Bologna, Italy}

\begin{document}

\maketitle

\begin{abstract}
We study the effect of a single, instantaneous starburst on the
dynamical and chemical evolution of a gas-rich dwarf galaxy, whose
potential well is dominated by a dark matter halo.  We follow the
dynamical and chemical evolution of the ISM by means of an improved
2-D hydrodynamical code coupled with detailed chemical yields
originating from type II SNe, type Ia SNe and single low and
intermediate mass stars (IMS). In particular we follow the evolution
of the abundances of H, He, C, N, O, Mg, Si and Fe.  We find that for
a galaxy resembling IZw18, a galactic wind develops as a consequence
of the starburst and it carries out of the galaxy mostly the
metal-enriched gas. In addition, we find that different metals are
lost differentially in the sense that the elements produced by type Ia
SNe are more efficiently lost than others. As a consequence of that we
predict larger [$\alpha$/Fe] ratios for the gas inside the galaxy than
for the gas leaving the galaxy. A comparison of our predicted
abundances of C, N, O and Si in the case of a burst occurring in a
primordial gas shows a very good agreement with the observed
abundances in IZw18 as long as the burst has an age of $\sim 31$ Myr
and IMS produce some primary nitrogen.  However, we cannot exclude
that a previous burst of star formation had occurred in IZw18
especially if the preenrichment produced by the older burst was lower
than $Z=0.01$ Z$_{\odot}$. Finally, at variance with previous studies,
we find that most of the metals reside in the cold gas phase already
after few Myr. This result is mainly due to the assumed low SNII
heating efficiency, and justifies the generally adopted homogeneous
and instantaneous mixing of gas in chemical evolution models.

\end{abstract}


\section{Introduction}

Dwarf irregular galaxies (DIG) are playing an increasingly central
role in understanding galaxy evolution. This kind of galaxies
generally has a low metallicity (from 0.5 Z$_{\odot}$ to
0.02 Z$_{\odot}$), a high gas content (up to $\sim$ 10 times the
stellar content) and their stellar populations appear to be mostly
young. All these features indicate that these galaxies are
poorly evolved objects.

Many gas-rich dwarf galaxies are known to be in a starburst phase, or
are believed to have experienced periods of intense star formation in
the recent past. These galaxies are generally called blue compact
dwarf (BCD) galaxies (Sandage \& Binggeli 1984). In general, dwarf
gas-rich galaxies, given their simple structures and small sizes, are
excellent laboratories to investigate the feedback of starbursts
on the interstellar medium (ISM), and to study their chemical
evolution. The aim is to reproduce the observed abundance ratios, to
trace their recent star formation history and to discover if these
galaxies could be the source of the intracluster gas (Gibson \&
Matteucci 1997).

Many authors have tried to connect late-type gas-rich (DIG and BCD)
and early-type gas-poor dwarf galaxies (dwarf ellipticals and dwarf
spheroidals) in an unified evolutionary scenario. The favourite theory
about ISM depletion in gas-rich dwarf galaxies is based on the
starburst-driven mass loss. The basis of this model, proposed by
Larson (1974) and then applied specifically to dwarfs by Vader (1986)
and Dekel \& Silk (1986), is that the ISM is blown out of the galaxy
by the energetic events associated with the star formation (stellar
winds and supernovae). The well-known correlation between mass and
metallicity found for both late-type and early-type dwarf galaxies
(Skillman, Kennicut \& Hodge 1989) is a natural result of the
increasing inability of massive galaxies to retain the heavy
elements produced in each stellar generation. At the present time is
not yet clear if galactic winds are really the key point for
understanding the formation and evolution of dwarf galaxies (see
Skillman \& Bender 1995 and Skillman 1997 for critical reviews about
this point), but they certainly play an important role, regulating the
mass, metal enrichment and energy balance of the ISM.

Observational evidences in support of the presence of outflows have
been found recently in a lot of gas-rich dwarf galaxies, like NGC1705
(Meurer et al. 1992), NGC1569 (Israel 1988), IZw18 (Martin 1996)
and many others. In their search for outflows in dwarf galaxies,
Marlowe et al. (1995) pointed out that this kind of phenomena is
relatively frequent in centrally star-forming galaxies. Again, they
note a preferencial direction of propagation along the galaxy minor
axis. In spite of these observational evidences, it is often difficult
to estabilish if the gas will leave definitively the parent galaxy. In
order to understand the final fate of both the swept-up gas and the
metals ejected during the starburst and to study possible links
between early and late-type dwarfs, numerical simulations are needed.
 
There are a lot of recent hydrodynamical simulations concerning the
behaviour of the ISM and the metals ejected by massive stars after
a starburst.  These simulations generally agree on the fact that
galactic winds are not so effective in removing the ISM from dwarf
galaxies, but disagree on the final fate of the metal-enriched gas
ejected by massive stars. Many authors (D'Ercole \& Brighenti 1999,
hereafter DB; MacLow \& Ferrara 1999, hereafter MF; De Young \&
Heckman 1994; De Young \& Gallagher 1990) have found that galactic
winds are able to eject most of the metal-enriched gas, preserving a
significant fraction of the original ISM. Other authors (Silich \&
Tenorio-Tagle 1998; Tenorio-Tagle 1996) have suggested that the
metal-rich material is hardly lost from the galaxies, since it is at
first trapped in the extended haloes and then accreted back on to the
galaxy.

However, all these models consider only the effects of stellar winds and
SNII explosions on the dynamics of the ISM. In this paper we present
models which take into account also the energetic contribution and the
feedback from intermediate-mass stars and SNeIa, using the most
up-to-date supernova rates. The effect of SNIa explosions is certainly
fundamental for the late dynamical evolution of the ISM (up to $\sim$ 500
Myr after the burst), even if their number is small.

There is an extensive literature about the chemical evolution of starburst
and blue compact dwarf galaxies (see e.g. Matteucci \& Chiosi 1983;
Matteucci \& Tosi 1985; Olofsson 1995). Pilyugin (1992, 1993) and
Marconi et al. (1994) suggested the idea that the spread in the
chemical properties of these galaxies, in particular the observed
spread in He/H vs. O/H and N/O vs. O/H, could be due to self-pollution
of H\,{\sc ii} regions coupled with `enriched' or `differential' galactic
winds.

Therefore it is interesting to test the differential wind hypothesis
with an hydrodynamical approach.  In our models we are able to follow
the evolution in space and time of the abundances of several
chemical elements (H, He, C, N, O, Mg, Si, Fe); in particular we
follow, with suitable tracers, the gas released by stars of different
initial mass. The chemical composition of each of these tracers is
obtained by adopting the nucleosynthesis prescriptions from various
authors (Woosley \& Weaver 1995, hereafter WW; Renzini \& Voli 1981,
hereafter RV; Nomoto, Thielemann \& Yokoi 1984, hereafter NTY).

In section 2 we describe the model and the assumptions adopted in our
simulations. The results are presented in section 3 and compared with
the observational constraints available for the BCG IZw18. A
discussion is presented in section 4 while some conclusions and future
improvements of the model are discussed in section 5.

\section{Assumptions and equations}

\subsection{The gravitational potential and the gas distribution}

It is convenient, for computational reasons, to model BCD galaxies.
In these galaxies, in fact, the starburst occurs near the optical
centre and the ISM structure is highly axisymmetric. In particular, we
will focus on the galaxy IZw18 which is a well-studied, very
metal-poor BCD galaxy.  IZw18 shows very blue colors ($U-B\,=-0.88$, Van
Zee et al.  1998), which are indicative of a dominating very young
stellar population, although one cannot exclude an underlying older
one (Aloisi et al. 1999).  Therefore, IZw18 is an excellent
candidate to compare with a single-burst model, although our model
cannot reproduce the real galaxy in detail.
 
Many ingredients play an important role in the dynamical evolution of
the ISM: the galactic structure (stellar component, gaseous component,
dark halo), the energy and mass injection rate of newly formed stars
and the size of the starburst region.

We model the ISM of IZw18 assuming a rotating gaseous component in
hydrostatic isothermal ($T_{\rm g}=10^3$ K) equilibrium with the
galactic potential and the centrifugal force. The potential well is
the sum of two components.  The first is given by a spherical,
quasi-isothermal dark halo truncated at a distance $r_{\rm {t h}}$, in
order to obtain a finite mass:
\begin{equation}
\rho_{\rm h}(r)=\rho_{\rm h 0}\biggl\lbrack{1+\biggl({r \over 
r_{\rm {c h}}}\biggr)^2}
\biggr\rbrack^{-1},
\end{equation}
\noindent
where $r=\sqrt{R^2+z^2}$ and $r_{\rm {c h}}$ is the core radius of the
dark component (we are using cylindrical coordinates). According to
values found in literature for the total mass of IZw18 (Lequeux \&
Viallefond 1980; Van Zee et al. 1998), the halo mass is assumed to
be $6.5 \times 10^8$ M$_{\odot}$. Since we do not take into account
the self gravity of the gas, in order to reproduce the oblate
distribution of gas inside IZw18 (Van Zee et al. 1998), we
introduce a fictitious `stellar' component described by an oblate
King stellar profile:
\begin{equation}
\rho_{\star}(R,z)=\rho_{\star 0}\biggl\lbrack{1+\biggl({R \over R_{\rm
c \star}} \biggr)^2+\biggl({z \over z_{\rm c
\star}}\biggr)^2}\biggr\rbrack^{-{3 \over 2}},
\end{equation} 
\noindent
where $R_{\rm c \star}$ and $z_{\rm c \star}$ are the core radius
along the $R$-axis and the $z$-axis respectively.  This profile is
truncated at the tidal radii $R_{\rm t \star}$ and $z_{\rm t \star}$,
in order to obtain a finite mass $M_{\star}=6\times 10^5$ M$_{\odot}$.
This structure is flattened along the $z$-axis and we assume $R_{\rm c
\star}/z_{\rm c \star}= R_{\rm t \star}/z_{\rm t \star}=5$ and $R_{\rm
t \star}/R_{\rm c \star}= z_{\rm t \star}/z_{\rm c \star}=4.29$. All
the structural parameters of our galactic model are summarized in
Table 1. The atomic number density of the neutral ISM is defined as
$n_{\rm g}={\rho\over 2 \mu m_{\rm H}}$, where $\rho$ is the ISM mass
density and $\mu=7/11$ is the mean mass per particle of the fully
ionized gas, assuming a primordial abundance.

Although this structure is rather flat, its potential is rounder. The
gas settled in such a potential assumes an oblate structure resembling
that of the ISM of IZw18 in a region $R\leq 1$ Kpc and $z\leq 730$
pc, which we call `galactic region'. We note however that the
elongation is also due to the assumed rotation of the gas which is
responsible for the flaring at large radii (see Fig. 1, upper panel).
Details of how to build such an equilibrium configuration can be found
in DB. The lower panel in Fig. 1 shows the resulting column density of
the ISM.

We ran several models varying the gas mass and the burst
luminosity. We discuss in detail three of them, M1, M2 and M3
(see Table 2). We also describe model MC, similar to M1, in
which heat conduction is allowed.

\par
\hbox{}
\begin{table*}
\caption[]{Galactic parameters}
\begin{flushleft}
\begin{tabular}{llllll}
\noalign{\smallskip}
\hline
\noalign{\smallskip}
$M_{\star}$(M$_{\odot}$) & $M_{\rm {halo}}$(M$_{\odot}$) & 
$R_{\rm c \star}(\pc)$ & $R_{\rm t \star}(\pc)$ & $r_{\rm {ch}}(\pc)$ & 
$r_{\rm {t h}}(\kpc)$\\
\noalign{\smallskip}
\hline\noalign{\smallskip}
$6\times 10^5$  & $6.5\times 10^8$ & 233 & 1000 & 
700 & 10\\  
\noalign{\smallskip}
\hline
\end{tabular}
\end{flushleft}
\end{table*}  

\par  
\hbox{}
\begin{table*}
\caption[]{ISM parameters}
\begin{flushleft}
\begin{tabular}{lllll}
\noalign{\smallskip}
\hline
\noalign{\smallskip}
Model &
$M_{\rm burst}$(M$_{\odot}$) & 
$M_{\rm g}$(M$_{\odot}$) & 
$n_{\rm g0}(\rm cm^{-3})$ & $E_{\rm b}({\rm erg})$\\
\noalign{\smallskip}
\hline\noalign{\smallskip}
M1 & $6\times 10^5$  & $1.7\times 10^7 $& $ 1.81$ & $1.5\times 10^{53}$\\
M2 & $3.6\times 10^5$  & $1.7\times 10^7$ & $1.81$ & $1.5\times 10^{53}$\\
M3 & $6\times 10^5$  & $4.6\times 10^6$ & $0.49$ & $4\times 10^{52}$\\
\noalign{\smallskip}
\hline
\end{tabular}

$M_{\rm g}$ is the ISM mass inside the galactic region defined in the
text, $M_{\rm {burst}}$ is the mass of the stars formed during the
burst and $n_{\rm g0}$ is the central atomic number density. 
$E_{\rm b}$ is the binding energy of the gas inside the galaxy.
\end{flushleft}
\end{table*}  

\begin{figure}
\centering
\vspace{-0.2cm}
\epsfig{file=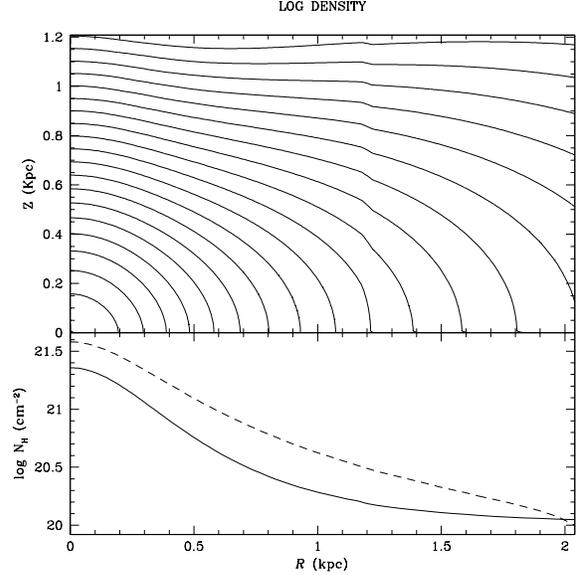, height=8cm,width=8cm}
\caption[]{\label{fig:fig 1} Upper panel: initial gas density
profiles. The density scale is logharitmic varies linearly between 
-33 and -24. Lower panel: column density of the initial ISM seen
edge-on (dashed line) and face-on (solid line).}
\end{figure}

\subsection{The equations}
To describe the evolution of the gas we solve the time-dependent, 
Eulerian equations of gasdynamics with source terms, that we write 
in the form:
\begin{equation}
{\partial\rho\over\partial t}+\nabla\cdot(\rho {\bmath{v}})=
\alpha\rho_*,
\end{equation}
\begin{equation}
{\partial\varrho^i\over\partial t}+\nabla\cdot(\varrho^i {\bmath{v}})=
\alpha^i\rho_*,
\end{equation}
\begin{equation}
{\partial {\bmath{m}}\over\partial t}+\nabla\cdot({\bmath{m}}\otimes 
{\bmath{v}})=\rho{\bmath{g}}-(\gamma-1)
\nabla\varepsilon + \alpha {\rho_*} {\bmath{v}}_*,
\end{equation}
\begin{equation}
{\partial\varepsilon\over\partial t}+\nabla\cdot(\varepsilon {\bmath{v}})=
-(\gamma-1)\varepsilon\nabla\cdot{\bmath{v}}-L+\alpha\rho_*
\biggl(\epsilon_0+{1\over 2}v^2\biggr),
\end{equation}
\noindent
where $\rho$, ${\bmath{m}}$ and $\varepsilon$ are the density of mass,
momentum and internal energy of the gas, respectively.  The parameter
$\gamma=5/3$ is the ratio of the specific heats, ${\bmath{g}}$ and
${\bmath{v}}$ are the gravitational acceleration and the fluid
velocity, respectively. The source terms on the r.h.s. of equations
(3)--(6) describe the injection of total mass and energy in the gas due
to the mass return and energy input from the stars.  In our
simulations the burst is located in the centre of the galaxy,
therefore both energy sources (SNeII and SNeIa) and mass return are
concentrated inside a small central sphere of $\sim$ 40 pc of
radius. We treat both sources as continuous, although the SNIa rate is
rather low ($\sim 1.6$ Myr$^{-1}$, see section 2.3.1). However, an a
posteriori analysis of our results, following Mac Low \& McCray
(1988), reveals that the continuous energy input assumption is still
valid during the SNIa stage. ${\cal V}$ being the volume of the burst,
$\rho_*=M_{\rm burst}/ {\cal V}$, where $M_{\rm burst}$ is the total
mass of stars formed during the burst (see next
section). ${\bmath{v}}_*$ is the circular velocity of these stars, and
$\alpha(t)=\alpha_*(t)+\alpha_{\rm SNII}(t)+\alpha_{\rm SNIa}(t)$ is
the sum of specific mass return rates from stars and SNe, respectively
(see next section). $\epsilon_0$ is the injection energy per unit mass
due to the stellar random motions and to SN explosions (see next
section).  Finally, $L=n_{\rm e}n_{\rm p}\Lambda(T)$ is the cooling
rate per unit volume, where for the cooling law $\Lambda(T)$ we follow
the approximation to the equilibrium cooling curve given by Mathews \&
Bregman (1978).

$\varrho^i$ represents the mass density of the $i$ element, and
$\alpha^i$ the specific mass return rate for the same element, with 
$\sum^\nn_{i=1} \alpha^i=\alpha$.
Eq. (4) represents a subsystem of $\nn$ equations which follow the
hydrodynamical evolution of $\nn$ different ejected elements (namely
H, He, C, N, O, Mg, Si and Fe). This enables us to calculate the
abundance of the ejecta relative to the pristine ISM.

To integrate numerically the eqs. (3)--(6) we used a 2-D hydrocode,
based on the original work of Bedogni \& D'Ercole (1986). We adopted a
non-uniform cylindrical axisymmetric grid whose meshes expand
geometrically.  The first zone is $\Delta R=\Delta z=5$ pc and the
size ratio between adiacent zones is 1.03.

\subsection{The starburst}
For the sake of simplicity we focus on a single, instantaneous,
starburst event located at the centre of the galaxy. The stars are
all born at the same time but they die and restore material into the
ISM according to their lifetimes.

Here we will describe the main assumptions about the initial mass function 
(IMF), stellar lifetimes and nucleosynthesis prescriptions adopted to
calculate $\alpha_*$ and $\alpha_{\rm SN}$.
\subsubsection{Mass return}
In order to obtain the number d$N$ of stars with initial masses in the
interval d$M$, we adopt the Salpeter (1955) initial mass function
(IMF) $\phi(M)={{\rm d}N\over {\rm d}M}$ assumed to be constant in
space and time:

\begin{equation}
\phi(M){\rm d}M=B M^{-(1+x)}{\rm d}M,
\end{equation}
\noindent
where $x=1.35$, and $B$ is the normalization constant obtained from:

\begin{equation}
\int^{40}_{0.1}{M \phi(M) {\rm d}M}=M_{\rm burst},
\end{equation}
\noindent
With $M_{\rm burst}=6\times 10^5$ M$_{\odot}$ we get a mass of $\sim 1.5
\times 10^{5}$ M$_{\odot}$ for the stars with masses larger than 2
M$_{\odot}$, in agreement with the estimate of the stellar content in
IZw18 by Mas-Hesse \& Kunth (1996).  Since the stellar yields are
calculated only for stellar masses not larger than 40 M$_{\odot}$
(WW), we adopt this value as an upper limit in eq. (8). Given the very
low number of stars more massive than this limit, the chemical and
dynamical evolution of the gas is not affected by this choice.

We assume that all the stars of initial mass between 8 and 40 solar
masses end their lifecycle as type II supernovae. The SNII rate is
defined as:

\begin{equation}
R_{\rm {SNII}}(t)=\phi(M)|\dot M|,
\end{equation}
\noindent
where $M$ represents the mass of the dying stars at the time $t$. 
The mass return rate from SNII is then given by:

\begin{equation}
\alpha_{\rm {SNII}}(t)=R_{\rm {SNII}}(t) \Delta M/M_{\rm burst}.
\end{equation}
\noindent
Here $\Delta M$ is the mass restored into the ISM by a star of initial
mass $M$, and is defined as $M-M_{\rm
rem}$, where $M_{\rm rem}$ is the mass of the stellar remnant.

In terms of single elements we have:

\begin{equation}
\alpha^{i}_{\rm {SNII}}(t)=R_{\rm {SNII}}(t) \Delta M_i/M_{\rm burst},
\end{equation}
\noindent
where $\Delta M_i$ is the mass restored by a star of mass 
$M$ in the form of the specific element $i$.

The specific mass return from stars with $M<8$ M$_{\odot}$ is given by:

\begin{equation}
\alpha_*(t)=\phi(M)|\dot M|\Delta M/M_{\rm burst},
\end{equation}
\noindent
and:

\begin{equation}
\alpha^{i}_*(t)=\phi(M)|\dot M|\Delta M_i/M_{\rm burst},
\end{equation}
\noindent
where, again, $M$ is the mass of the dying stars at the time $t$.

To calculate the time derivative of the mass in eqs. (9), (12) and (13) we
adopt the stellar lifetimes given by Padovani \& Matteucci (1993):

\begin{equation}
t(M)=\cases{1.2 M^{-1.85}+0.003\;{\rm Gyr} &if $M\geq 8$ M$_{\odot}$\cr
     10^{f(M)}
\;{\rm Gyr} &if $M<8$ M$_{\odot}$,\cr}
\end{equation}
\noindent
where $f(M)={{\bigl\lbrack 0.334-\sqrt{1.79-0.2232\times(7.764-\log(M))}
\bigr\rbrack\over 0.1116}}$.

To obtain the quantity $\Delta M$ appearing in eq. (10) and eq. (12)
we took into account the results of WW for massive stars ($M \ge10$
M$_{\odot}$) and RV for low and intermediate mass stars ($0.8 \le
M/{\rm M}_{\odot} \le 8$), which give the mass restored into the ISM
by the stars at the end of their lifetime.  For the range 8
M$_{\odot}\leq M\leq 10$ M$_{\odot}$ we have adopted suitable
interpolations between the previous two sets of data.

In WW the total ejected masses (processed and unprocessed) are given
for each chemical element.  In general, however, in nucleosynthesis
papers only the `yield' is given, namely the fraction in mass of a
given element $i$ which is newly formed and ejected by a star of
initial mass $M$, the quantity $P_{i \rm M}$.  In this case, the
ejected total masses are computed in the following way:

\begin{equation}
\Delta M_i=\Delta M X_i +MP_{i \rm M},
\end{equation}
\noindent
where $X_i$ is the original abundance of the element $i$ in the star.
This is the case of the yields of RV.

From the tables of WW (which contain also the products of explosive
nucleosynthesis) and RV we have derived several relations between the
initial stellar mass and the mass restored into the ISM in the form of
chemical elements for single stars of masses between 0.8 and
40 M$_{\odot}$, obtained by fitting the tabulated values with an
eighth degree polynomial. The results are shown in Figg. 2--4 for
different initial chemical compositions and different mixing lenght
parameters. This enables us to obtain the temporal behaviour of
$\alpha_{*}^i(t)$ and $\alpha_{\rm SNII}^i(t)$ for each element $i$.
The total mass ejection rates obtained by summing over all the
chemical elements are: $\alpha_{\rm SNII}(t)\propto t^{-0.27}$ and
$\alpha_*(t)\propto t^{-1.36}$ (see also Ciotti et al. 1991).

Finally, in analogy with eq. (10) we define the specific mass return
from the SNeIa as:
\begin{equation}
\alpha_{\rm {SNIa}}(t)=1.4R_{\rm {SNIa}}(t)/M_{\rm burst},
\end{equation}
\noindent
and:
\begin{equation}
\alpha^{i}_{\rm {SNIa}}(t)=R_{\rm {SNIa}}(t) \Delta M_i/M_{\rm burst},
\end{equation}
\noindent
where the mass ejected by each SNIa is assumed to be 1.4 M$_{\odot}$
(the Chandrasekhar mass).  According to the single degenerate model
(SD), SNe Ia are assumed to originate from C-O white dwarfs in binary
systems which explode after reaching the Chandrasekhar mass as a
consequence of mass transfer from a red giant companion. This kind of
supernova explosion occurs only after the death of stars of initial
mass less or equal than 8 M$_{\odot}$, which is $\sim$ 29 Myr after
the burst. $R_{\rm SNIa}(t)$ is given by the following formula,
obtained by the best-fitting of the SNIa rate computed in detail
numerically by the model of Bradamante et al. (1998), when applied to
the case of a single starburst:

\begin{equation}
R_{\rm {SNIa}}(t)=4.2
\times 10^{-9}\biggl({{t_9+1}\over 15}\biggr)^{-1.9}\;{\rm yr}^{-1},
\end{equation}
\noindent 
where $t_9$ is the time expressed in Gyr. 

It is worth noting that the SNIa rate in BCG is practically unknown
and therefore it is very difficult to choose the right fraction of
binary systems, in the IMF of such galaxies, of the type required to
originate a SNIa. The rate of eq. (18) corresponds to the rate of
Greggio \& Renzini (1983) for a starburst with a fraction of binary
systems $A=0.006$. This rate switches on somewhat more gradually than
in our approximation, reaching a maximum after $\sim$ 40 Myr (see Greggio \&
Renzini, fig. 1), but this difference has no consequences in the
dynamical evolution of our models. To show this, we ran a model (not
shown in this paper), up to $\sim$ 80 Myr, using the rate computed by
Greggio \& Renzini and the differences with the results shown in
section 3 were negligible.  With our assumed rate, SNeIa
contribute by $\sim 60$ per cent of the total iron production after 15 Gyr,
in agreement with predictions for the solar neighbourhood (Matteucci
\& Greggio 1986).

\begin{figure*}
\centering
\vspace{0.1cm}
\epsfig{file=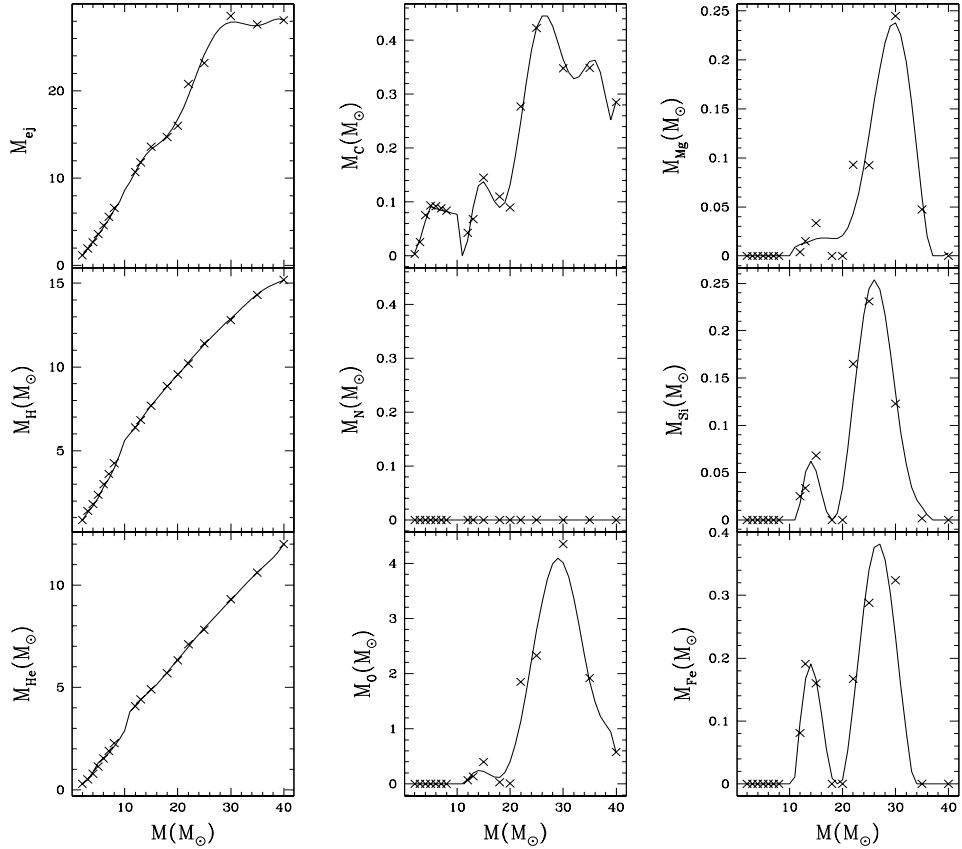,height=12.8cm,width=13.8cm}
\caption[]{\label{fig:fig 2} Yields from massive and intermediate stars 
(data taken from WW and RV yields) together with our fits (eighth degree 
polynomial best-fittings) for the case A. 
$M_{\rm {ej}}$ is the total 
mass of gas ejected by the star and corresponds to $\Delta M$ 
as defined in the text.}
\end{figure*}

\begin{figure*}
\centering
\vspace{0.1cm}
\epsfig{file=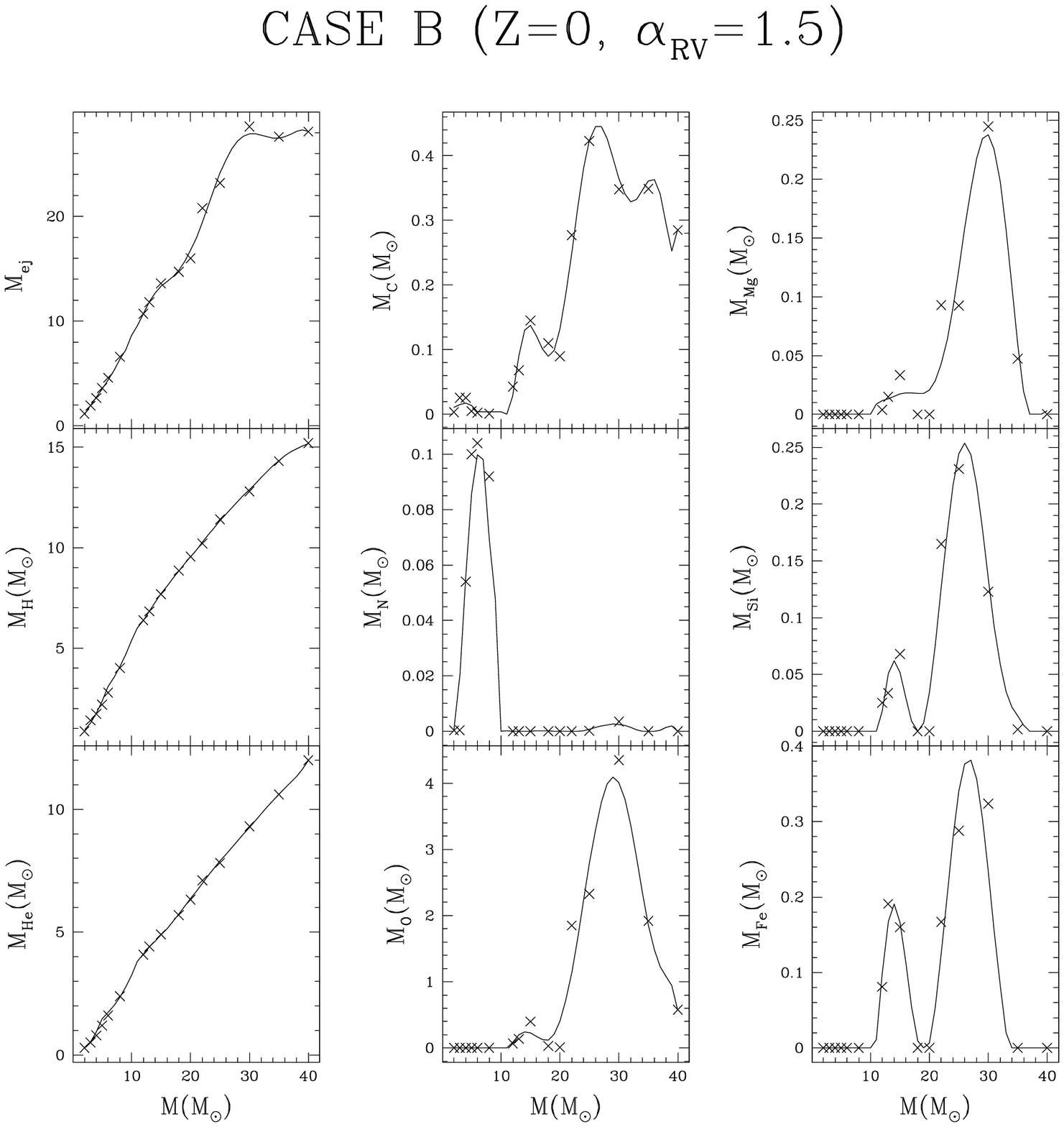,height=12.8cm,width=13.8cm}
\caption[]{\label{fig:fig 3} Yields from massive and intermediate stars 
for the case B.}
\end{figure*}

\begin{figure*}
\centering
\vspace{0.1cm}
\epsfig{file=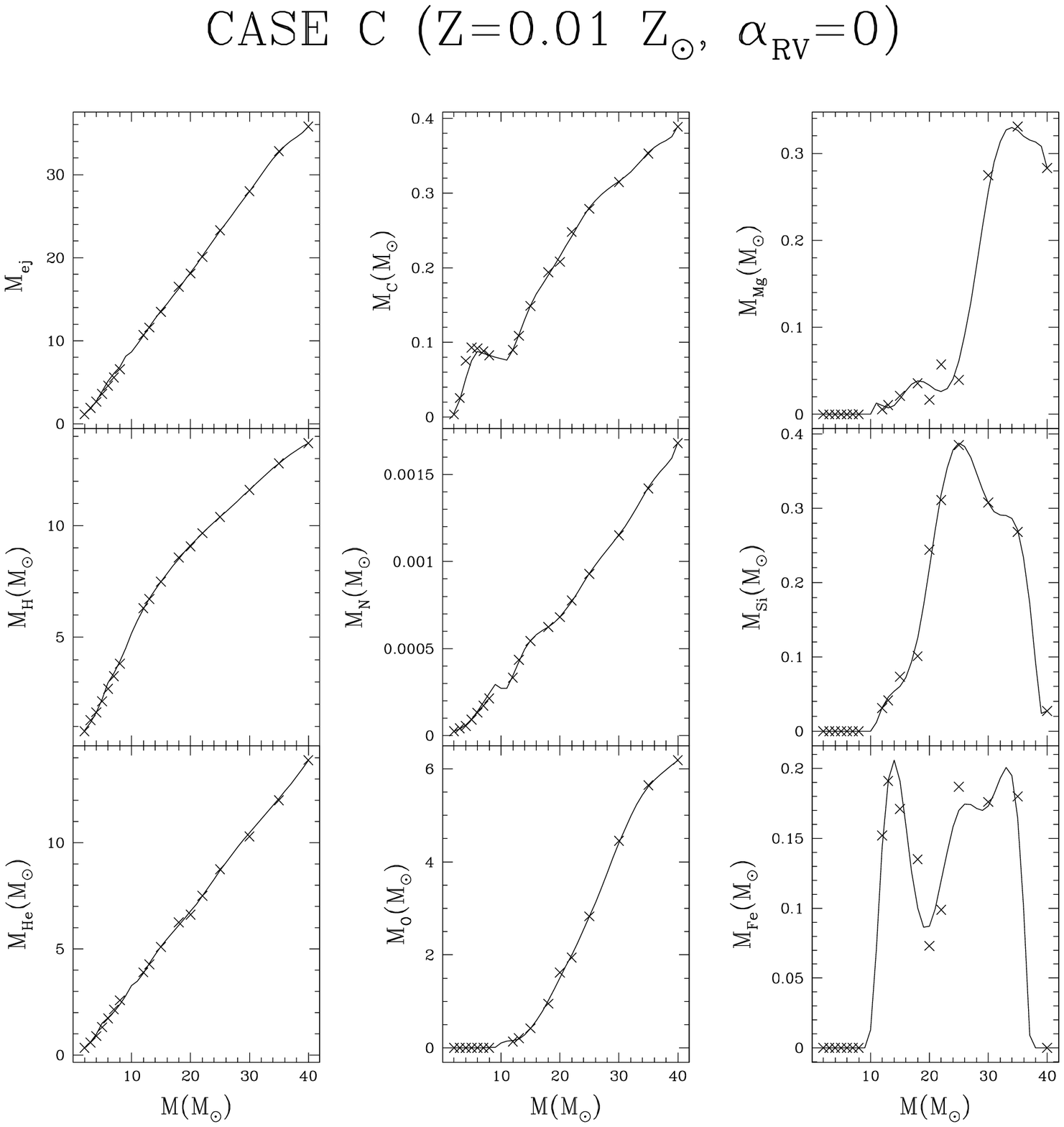,height=12.8cm,width=13.8cm}
\caption[]{\label{fig:fig 4} Yields from massive and intermediate stars 
for the case C.}
\end{figure*}

In summary, stars in different mass ranges contribute to galactic 
enrichment in a different way:

\begin{enumerate}
\item For low and intermediate stars (0.8 M$_{\odot}\leq M\leq 8$ 
M$_{\odot}$) we have used the RV nucleosynthesis calculations for a
value of the mass loss parameter $\eta=0.33$ (Reimers 1975) and the
mixing lenght $\alpha_{\rm {RV}}=0$ and $\alpha_{\rm {RV}}=1.5$. The
initial chemical composition is either $Z=0$ or $Z=1/100$ Z$_{\odot}$.
These stars mainly produce He, C, N and s-process elements (not
considered here).  In particular, N is a `secondary' element, namely
produced from the original C and O present in the star at
birth. Therefore, for zero metallicity initial chemical composition no
N would be produced.  However, there is the possibility of producing N
in a `primary' way, namely starting from the C and O newly formed in
the star.  This is the case of the IMS which can produce primary N during
the third dredge-up episode in conjunction with the hot-bottom
burning, during the thermal-pulsing phase occurring when these stars
are on the asymptotic giant branch (AGB) (case $\alpha_{RV}=1.5$ of
RV).  Moreover, massive stars can also produce primary N, as suggested
by Matteucci (1986).  In the nucleosynthesis prescriptions of WW there
is some primary N from massive stars but is negligible.

\item 
For massive stars ($M>10$ M$_{\odot}$) we have adopted the case B in
the WW nucleosynthesis results, focusing our attention on the models
with $Z=0$ and $Z=1/100$ Z$_{\odot}$.  These stars are responsible for the
production of the $\alpha$-elements (O, Mg and Si) and for part of the
iron.  The stars in the mass range 8 M$_{\odot}\leq M\leq
10$ M$_{\odot}$ produce mainly He and some C, N and O.

\item 
For type Ia SNe we have followed the results of NTY adopting their model
W7. In this model, every type Ia SN restores into the ISM $\sim$ 1.4
M$_{\odot}$ of gas.  Most of this gas is ejected in the
form of Fe ($\sim$ 0.6 M$_{\odot}$) and the rest is in the form of
elements from C to Si.
\end{enumerate}

In Table 3 we present a brief summary of the nucleosynthesis
prescriptions adopted in our models. Note that each of these cases can
be adopted for the three different hydrodynamical models. Thus, for
instance, hereafter with model M1B we intend the hydrodinamical model
M1 with the chemical option B.

\par
\hbox{}
\begin{table}
\caption[]{Nucleosynthesis prescriptions}
\begin{flushleft}
\begin{tabular}{lll}
\noalign{\smallskip}
\hline
\noalign{\smallskip}
Case & $Z$ & $\alpha_{\rm {RV}}$ \\
\noalign{\smallskip}
\hline\noalign{\smallskip}
A & 0 & 0 \\
B & 0 & 1.5 \\
C & 0.01 Z$_{\odot}$ & 0 \\
D & 0.01 Z$_{\odot}$ & 1.5\\
\noalign{\smallskip}
\hline
\end{tabular}
\end{flushleft}
$Z$ is the abundance of the unprocessed gas. If $Z=0$, we assume
$X=0.77$, $Y=0.23$.  The solar abundances adopted are taken from 
Anders \& Grevesse (1989).
\end{table}  
\noindent

\subsubsection{Energy input}
The energy input into the ISM due to the stellar activity is taken
into account in eq. (6) through the term $\epsilon_0=3kT_0/2\mu$,
where $k$ is the Boltzmann constant. The injection temperature can be
written as:
\begin{equation}
T_0=(\alpha_*T_*+\eta_{\rm II}\alpha_{\rm SNII}T_{\rm II}+\eta_{\rm
Ia}\alpha_{\rm SNIa}T_{\rm Ia})/\alpha,
\end{equation}
\noindent
where $kT_*$, $kT_{\rm II}$ and $kT_{\rm Ia}$ are the energy per unit
mass in the ejecta of stars, SNeIa and SNeII, respectively (see
e.g. Loewenstein \& Mathews 1987 for more details). $\eta_{\rm II}$
and $\eta_{\rm Ia}$ represent the efficiency with which the energy of
the stellar explosions is transferred into the ISM for SNeII and
SNeIa, respectively. We assume that 10$^{51}$ erg of mechanical energy
are produced during the explosion of both types of supernova. However,
we assume $\eta_{\rm II}=0.03$; only 3 per cent of the energy explosion is
available to thermalize the ISM, while the rest is radiated away.
This prescription is taken from the work of Bradamante et al. (1998),
who studied in detail the chemical evolution of blue compact
galaxies. Actually, some debate is present in literature about the
efficiency of the SNII in heating the ISM in starbursts. We thus run
also a model with $\eta_{\rm II}=1$ which however is not succesful in
describing IZw18, as discussed at the end of section 3.1.5.

For the SNIa explosions, instead, we assume
$\eta_{\rm Ia}=1$ because the SNRs expansion occur in a medium already
heated and diluted by the previous activity of SNII.

It is worthwhile to note that we neglected the energetic contribution
of stellar winds from massive stars, according to the results of
Bradamante et al. (1998) who showed that the injected stellar wind
energy is negligible relative to the SN energy in this
kind of galaxies and to the results of Leitherer et al. (1999). This
is mostly a consequence of the low initial metallicity adopted ($Z=0$ or
$Z=0.01$ Z$_{\odot}$), because the mass loss from stars strongly depends
on their metallicity (see e.g. Portinari et al. 1998 and references
therein).

\section{Simulations}

\subsection{Dynamical results}

Our reference model is M1. In addition, we run two other models, M2
and M3, which have the same potential well (see Table 2). In model M2
the mass $M_{\rm burst}$ of gas turned into stars is halved in
comparison with M1. In this case two competing effects are
expected. On one hand, half of the metals are produced and the
resulting increase in the metallicity of the ISM is expected to be
lower; on the other hand, the galactic wind luminosity powered by SNe
is also halved, stellar ejecta are expelled less effectively from the
galaxy and the enrichment of the galactic ISM tends to be higher. Two
competing tendencies are also present in model M3 which has nearly one
fourth of the ISM mass in comparison with M1. In this case the stellar
ejecta mixes with less gas and the metallicity of this gas is thus
expected to become higher than in M1, but in this case the wind is
favoured and less metals are retained by the galaxy, resulting in a
lower chemical enrichment. We also show model MC, similar to M1, to
study the action of heat conduction.

In order to discuss the hydrodynamical behaviour of the gas we recall
briefly a few results about bubble expansion in stratified media (Koo
\& McKee 1992, and references therein). The freely expanding wind
produced by the starburst interacts supersonically with the
unperturbed ISM thus creating a classical bubble (Weaver et al. 1977)
structure in which two shocks are present. The external one propagates
through the ISM giving rise to an expanding cold and dense shell,
while the inner one thermalizes the impinging wind producing the hot,
rarefied gas of the bubble interior. The shocked starburst wind and
the shocked ISM are separated by a contact discontinuity.  The density
gradient of the unperturbed ISM being much steeper along the $z$ axis,
the expansion of the outer shell occurs faster along this direction. A
bubble powered by a constant wind with velocity $V_{\rm w}$, mass loss
rate $\dot M$ and mechanical luminosity $L_{\rm w}=0.5\dot MV^2_{\rm
w}$ is able to break out from a gaseous disc if $L_{\rm w}>3L_{\rm
b}$, where the critical luminosity is $\lb=17.9\rho_0 H_{\rm
eff}^2C_0^3$. $C_0$ is the sound speed of the unperturbed medium and
$\rho_0$ is its central density.  $H_{\rm eff}$ is the effective scale
length of the ISM distribution in the vertical ($z$) direction and is
defined as:
\begin{equation}
H_{\rm eff}={1 \over\rho_0} \int_0^{\infty} \rho dz.
\end{equation}
\noindent
In our models $H_{\rm eff}\sim 300$ pc. If the wind luminosity is
larger than $L_{\rm n}\sim 0.35 m L_{\rm b}$, where $m=V_{\rm w}/C_0$,
the wind blows directly out of the planar medium, at least in
directions close to the axis. If, instead, $L_{\rm w} \simlt L_{\rm
n}$ the formation of a jet is possible, in which the wind is shocked
and then accelerated again to supersonic speeds through a sort of de
Laval nozzle created by the shocked ambient medium. Kelvin-Helmholtz
instabilities tend to distort the nozzle, and stable jets can exist
only for $\beta=C_{\rm h}/C_{0}\simlt 30$, where $C_{\rm h}$ is the
cavity sound speed (Smith et al. 1983).

\subsubsection {Model M1}

For model M1 $L_{\rm b}=2.8\times 10^{36}$ erg s$^{-1}$. Actually,
$\lw$ is not constant in our simulations. However, as a representative
value, for the wind powered by SNeII we have $L_{\rm w}\sim 2\times
10^{38}$ erg s$^{-1}$ (cf. Fig. 9) and $m\sim 300$, thus the bubble
carved by this wind is able to break out. Note that $\lw < L_{\rm
n}$, so that a jet-like structure is expected. As shown in Fig. 5, a
jet actually propagates with a shock velocity $V_{\rm s}\sim 3\times
10^6$ cm s$^{-1}$ after 30 Myr.  This figure also shows that the
bubble shell on the simmetry plane has reached the maximum allowed
value (Koo \& McKee 1992) $R_{\max}=0.72 H_{\rm eff}(\lw /\lb)^{1/6}
\sim 440$ pc.

The SNII wind lasts a relatively short time (29 Myr), and is then
replaced by a weaker SNIa wind with $L_{\rm w}\sim 2\times 10^{37}$
erg s$^{-1}$ (cf. Fig. 9) and $m\sim 200$. The existing jet cannot be
sustained by this wind and is inhibited before breaking out. The
bubble as a whole stops to grow, and the incoming shocked wind pushes
a large fraction of the hot SNII ejecta against the dense and cold
cavity walls. Thus most of these ejecta would be located close to the
cavity edge. The thermal evolution of these ejecta is difficult to
asses. Given the spread of contact discontinuities due to the
numerical diffusion, the ejecta partially mixes with the cold wall of
the cavity, so that a large fraction of these metals cools off
(cf. Fig. 14). Actually, as discussed by DB, several physical
processes, such as thermal conduction and turbulent mixing, produce a
similar effect. In section 3.2 we consider explicitly heat conduction,
but neglect the turbulent mixing (Breitschwerdt \& Kahn 1988, Kahn \&
Breitschwerdt 1989, Begelman \& Fabian 1990, Slavin, Shull \& Begelman
1993) which is very complex and nearly impossible (and probably even
meaningless out of a fully 3D simulation) to implement into the code.

The bubble inflated by the SNIa wind is likely to break out through a
nozzle. We note, however, that the gas distribution in front of the
bubble along the $z$ direction is modified by the expansion of the
outer shock generated by the previous SNII activity; although the
powerful SNII wind is ceased, this shock continues to expand with
increasing velocities ($V_{\rm s}\sim 10^7$ cm s$^{-1}$ at $t=342$
Myr) because of the steep gradient of the unperturbed ISM density
profile. At these shock velocities, the post-shock gas cools quickly
and its temperature is $T=10^3$ K (the minimum allowed in our
computations) everywhere with the exception of a `rim' behind the
shock, where $T\sim 3\times 10^5$ K.  The density gradient of the
upwind gas experienced by the SNIa bubble is shallower, and the break
out is contrasted.  This can be seen in Fig. 5, where the hot bubble
is shown to grow very little up to $t\sim 300$ Myr.

The gas in the bubble radiates inefficently because of its low density
($n\sim 3\times 10^{-4}$ cm$^{-3}$) and its temperature is $\sim
2\times 10^6$ K.  Shears are present at the contact surface between
hot and cold gas, which is thus Kelvin-Helmholtz unstable. For this
reason the cavity is irregularly shaped at this stage.

Subsequently, since the surrounding cold gas is in expansion, the hot
gas is finally able to break out carving a long tunnel. This tunnel
has the de Laval nozzle structure, with the transverse section
increasing with $z$.  We note the presence of Kelvin-Helmholtz
instabilities at the wall of the nozzle. This is due to the fact that
$\beta\sim 30$, so the nozzle is only marginally stable.

The shocked wind is accelerated again to supersonic speeds through
this nozzle (velocities $V\sim 4\times 10^7$ cm s$^{-1}$ and mach
numbers ${\cal M}\sim 8$). When the jet is well developed ($t=342$ Myr
in Fig. 5b), the minimum radius of the nozzle is $R_{\rm n}\sim 100$ pc.

The acceleration of the cold shell in front of the jet causes it to be
disrupted by the Rayleigh-Taylor instabilities, and the hot gas leaks
out. Contrary to the previous works where the SNIa activity was not
considered, at these late times the central galactic region is not yet
replenished by the cold surrounding gas. Taking into account equation
(18), we stress that this will happen after $\sim 2$ Gyr, when $\lw
=2\,\lb$.

In Fig. 9 we have plotted mass, energy and luminosity budget inside
the galaxy. From the central panel of this figure we note that the
energy of SNe never becomes larger than the binding energy, although
some gas is definitively lost from the galaxy, as it is apparent from
the numerical simulation. This indicates that ballistic arguments
cannot be adopted properly to calculate ejection efficiencies. In fact,
an element of fluid can acquire energy at the expense of the rest of the gas
through opportune pressure gradients, thus increasing its velocity
beyond the escape velocity.

The thermal energy shows in particular two drops at $t\sim$ 30 Myr and
at $t\sim$ 160 Myr. The second drop reflects a decrease in the hot gas
content, while the first one cannot be associated at any particular
hot gas loss. In fact this drop coincides with the discontinuity
SNII/SNIa, when the specific energy injection falls by a factor $\sim$
10. Thus the thermal content of the bubble decreases via radiative
cooling, although its temperature remains larger than $2\times\,10^4$
K, which is the threshold adopted to define hot regions in the upper
panel of Fig. 9. The fall-off at $t\sim$ 160 Myr is instead due to the
presence of large eddies which move part of the hot gas outside the
galaxy. We finally point out that at the beginning of the SNII
activity the X-ray emission is absent (see lower panel of Fig. 9)
because the energy injection is not able to rise the cavity
temperature over $T=7\times 10^5$ K, which is the threshold adopted to
define the X-ray emitting gas.

\begin{figure*}
\centering
\vspace{-6.5cm}
\vskip 30cm
\vspace{-0.5cm}
\caption{Density contours and velocity fields for the model M1 at
different epochs. The density scale (logarithmic) is given in the
strip on top of the figure. In order to avoid confusion, we draw only
velocities with values greater than 1/10 of the maximum value. This is
true also for Figg. 6, 7 and 8.}
\end{figure*}

\begin{figure*}
\centering
\vspace{-7cm}
\vskip 30cm
\vspace{-1.5cm}
\caption{As Fig. 5, but for model M2.}
\end{figure*}

\begin{figure*}
\vspace{-5cm}
\centering
\vskip 30cm
\vspace{-3.5cm}
\caption{As Fig. 5, but for the model M3.}
\end{figure*}

\begin{figure*}
\vspace{-5cm}
\centering
\vskip 30cm
\vspace{-3.5cm}
\caption{As Fig. 5, but for the model MC.}
\end{figure*}

\subsubsection {Model M2}

In this model, $\lb$ is the same as in M1, but $\lw$ is a factor of
0.6 lower. The dynamical evolution is rather similar to that of the
previous model. Obviously the bubble is smaller at the end of the SNII
activity. Quite surprisingly, however, the nozzle carved by the SNeIa
breaks out earlier than in M1. Note that in this case $\lw / \lb <
10$, and the nozzle is stable and well shaped. Actually, this
condition is equivalent to the condition $\beta\simlt 30$ expressed
above (Koo \& McKee 1992) and, for this model, we find $\beta\sim 25$.
Kelvin-Helmholtz instabilities are present, but they are carried away
by the flow before they can grow significantly.  Less energy is
dissipated by the turbulence and is more easily channelled through the
nozzle.

\subsubsection {Model M3}

Because of the lower ISM density, we have $\lb = 7.5\times 10^{35}$ erg
s$^{-1}$ for this model, so that $\lw > \ln$ during the SNII stage. In
this case the galactic wind breaks out rather vigorously, and the
evolution of the gas is similar to that found in other theoretical
works (Suchkov et al. 1994, De Young \& Heckman 1994, MF, DB). A
prominent lobe is formed, which is similar to that described by Mac
Low, McCray \& Norman (1989) in their fig. 2. The shell accelerates
and becomes Rayleigh-Taylor unstable, expelling the hot interior. The
breakup occurs at the polar cap, where the ISM has the lower density
and pressure. The hot gas which blows out of the bubble produces the
jet-like structure visible in Fig. 7a. Note that the outer shock
initially is rather slow ($V_{\rm s}\sim 5\times 10^6$ cm s$^{-1}$),
and then accelerates somewhat up to $V_{\rm s}=1.6\times 10^7$ cm
s$^{-1}$. Thus, also in this model the shocked gas never reaches high
temperatures and most of the lobe volume is cold.

Fig. 7b shows the late dynamical evolution of the gas. The gas flow
expands along a conical configuration, inside a solid angle which
remains constant during all the simulation. The ISM outside this
funnel remains substantially unperturbed. Actually, the aperture of
the cone in our model is evidently dictated by the assumed structure
of the ISM and could be not realistic. However, given that no falling
back or fountain is expected by the expelled material which is kept in
expansion by the SNeIa, the final chemical characteristics of the gas
inside the galaxy are not affected by the exact structure of the
`chimney'.

Concerning the distribution along the $z$ direction, the lobe of
outflowing gas can be grossly divided in three regions. The most
external, far from the galaxy, is bounded by the outer shock and hosts
mainly shocked, low-density external medium. The inner region is filled
with gas ejected by SNeIa and low-mass stars. Between these two
regions there is the gas of the `first' bubble (where the ejecta of
SNII are present), which is quickly cooled to low temperatures. At
$t\sim 150$ Myr, the shell of the small bubble breaks, like the first
one, and the hot gas flows forward into the lobe rising its
temperature up to 10$^6$ K.

Just before breaking out, the superbubble reaches the edge of the
galaxy even along the $R$-axis, pushing out all the unprocessed gas
present, and almost all the galaxy is covered by the hot cavity. As
the breakout occurs, the bubble shell shrinks slightly and part of it
comes back into the galactic region producing the rise of the mass of
hydrogen and the other elements (cf. Fig. 11). This happens in
coincidence with the pressure decrease in the SNIa bubble following
the rupture of the unstable shell, as discussed above.

\subsubsection{Model MC}

In Fig. 8 we show model MC, identical to model M1 but with the heat
conduction activated. In order to take into account the thermal
conduction, we solve the heat transport term through the
Crank-Nicholson method which is unconditionally stable and second
order accurate (see DB for more details). In this model the cavity is
less extended than in M1 because of the increased radiative losses due
to the evaporation front. During the SNII stage the bubble never
extends beyond $H_{\rm eff}$. Thus the ``nose'' present in M1 does not
develop, and the bubble has a more round aspect (Fig. 8a). Thermal
conduction smooths temperature inhomogeneities and the cavity is more
regularly shaped also at later times (see Fig. 8b). Less energy is
dissipated through eddies, and the final break out is
slightly anticipated compared to model M1. The resulting outflow is
stable and well-shaped. The fraction of cold ejecta does not change
substantially compared to model M1 (see section 3.2 for a discussion
about this point).

\begin{figure}
\centering
\vspace{-0.2cm}
\epsfig{file=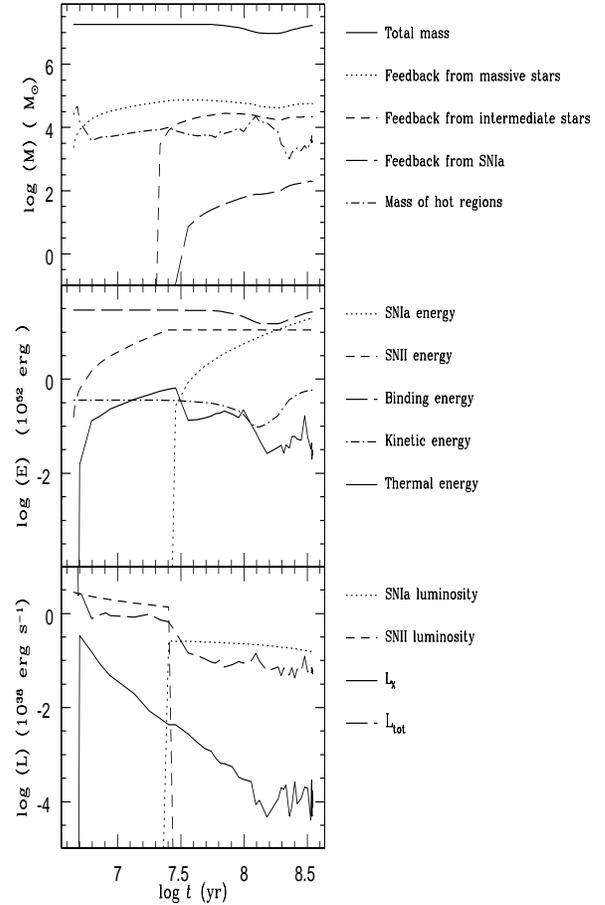,height=12.4cm,width=8.5cm}
\caption[]{\label{fig:fig 9} Energy, luminosity and mass budget inside
the galaxy for model M1.  Hot regions are defined as the regions where
$T>2\times 10^4 K$.  $L_X$ indicates the emission from gas with
$T>7\times 10^5 K$ (emission in the X-ray band), while $L_{\rm tot}$
is the total emission.}
\end{figure}

\subsubsection{ISM ejection efficiencies}

It is useful to define the efficiency of gas removal from the galaxy.
This efficiency cannot be unambigously defined as in the case of
ballistic motions because of dissipative effects which may play a very
important role (DB). For this reason we simply define the efficiency
$f_{\rm ISM}$ by dividing the mass of the gas which has left the
galaxy by the total mass of the ISM.  In a similar way we calculate the
efficiency in the ejection of material from SNII ($f_{\rm {SNII}}$),
intermediate-mass stars ($f_{\rm {IMS}}$) and SNIa ($f_{\rm {SNIa}}$).

The upper panels of Fig. 11 shows the masses of the different elements
removed from the galaxy as functions of time. The relative proportions
between masses of different elements is essentially that expected by a
Salpeter IMF and is not substantially affected by selective dynamical
losses. For the model M1B, we point out that the mass of metals lost
from the galaxy declines after reaching a maximum at $t\sim$ 200
Myr. This maximum is mirrored by a minimum in the masses of metals
inside the galaxy. This behaviour is due to the fact that, at this
time, the large hot blob visible in Fig. 5b (second panel) extends
over the galactic edge ($R$ direction) thus inducing a large loss of
ISM (mostly H and He).  We calculate the efficiency at $t\sim$ 200 Myr
and at the end of the simulation ($t\sim$ 375 Myr), obtaining $(f_{\rm
ISM},f_{\rm SNII},f_{\rm SNIa},f_{\rm IMS})$=(0.43,0.38,0.25,0.38) and
$(f_{\rm ISM},f_{\rm SNII},f_{\rm SNIa},f_{\rm
IMS})$=(0.07,0.17,0.20,0.26), respectively. Thus, at $t\sim$ 200 Myr
the products of the SNII and IMS have been ejected more easily than
the products of SNeIa. This is of course due to the fact that SNIa
material is located in a region closer to the galactic centre. After
the break up all the efficiencies decrease, and in particular $f_{\rm
SNII}$ shows the greatest reduction. In fact, the SNII ejecta inside
the galaxy are `incorporated' into the cold, dense shell of the cavity
and do not experience any substantial further dynamical evolution (see
discussion in section 3.1.1 and Fig. 5b); at the break out the bubble
shrinks and its walls recede entirely in the galaxy, increasing its
content of SNII ejecta.  Contrary to the expectations, the higher
efficiency is given by $f_{\rm IMS}$ instead of $f_{\rm SNIa}$. This
is due to strongly unsteady behaviours of the nozzle wall.

In model M2, where a nearly steady flow is obtained, we actually have
$f_{\rm SNII}<f_{\rm IMS}<f_{\rm SNIa}$.  At $t\sim$ 300 Myr it is
$(f_{\rm ISM},f_{\rm SNII},f_{\rm SNIa},f_{\rm IMS})$
=(0,0.06,0.32,0.12) in this model. For this model the difference in
efficiencies between the total gas and metals is particulary striking
and indicates that the differential galactic wind assumption, adopted
in several one-zone chemical models, is a natural outcome in this
scenario. In particular we note that at late times the galaxy is
almost completely replenished by gas. In fact, as apparent in Fig. 6b,
the nozzle has a rather small section ($R_{\rm n}\sim 85$ pc) and the
volume of the cavity is negligible in comparison with the galactic
volume. The more regular hydrodynamical behaviour reflects also in the
more regular temporal trend of the ejected masses. Note that, as
expected, in M2 the galactic wind starts later in comparison with M1.

Model M3 predicts of course the maximum amount of metals lost and is
also the first in which the break out occurs. The striking minimum in
the metal contents of the galaxy occurring at $t\sim$ 130 Myr is due
to the dynamical behaviour of the buble shell, as discussed above. 
The efficiencies for the model M3 at the end of the simulation ($t\sim$ 
470 Myr) are $(f_{\rm ISM},f_{\rm SNII},f_{\rm SNIa},f_{\rm IMS})$
=(0.77,0.85,0.97,0.87).

Finally, we mention that, as outlined in section
2.3.2, we run a model (not shown here) similar to M1, but with
$\eta_{\rm II}=1$. At the end of the SNII stage, the galaxy is almost
devoided of gas (a part for the tenuous galactic wind). The time-scale
for the galactic replenishment (the SNIa wind cannot preserve an empty
region so large) is at least $R_{\rm t*}/C_0 \sim$ 200 Myr. Actually,
due to the retarding effect of the centrifugal force, in our
simulation most of the galaxy is replenished after $\sim$ 450 Myr. The
age of the (last) burst occurred in IZw18 is estimated to be $\simlt$
27 Myr (see below), thus its actual content of gas rules out the
possibility of an high $\eta_{\rm II}$. Note that our assumption that
all energy injection occurs in the central region leads to the most
effective gas removal for a given luminosity (Strickland \& Stevens
1999). Models with a more realistic burst diffuse across the galaxy
will be presented in a next paper.

\subsection{Instantaneous versus delayed mixing}

In the previous models a large fraction of the stellar ejecta cools
quite soon and a rapid mixing is expected given the relatively short
diffusion time at these temperatures (for a comparison of the
diffusion times in the different ISM phases, see for example
Tenorio-Tagle 1996). This is an important point in view of the confrontation of
the results of our models with the $observable$ abundances of IZw18
(cf. section 4.2), and deserves some discussion.

Rieschick
\& Hensler (2000), for instance, presented a chemodynamical model of
the ISM of a dwarf galaxy in which the metal enrichment undergoes a
cycle lasting almost 1 Gyr. This model is based on the scenario
described in Tenorio-Tagle (1996). In this scenario the break out is
inhibited and the SNII ejecta (SNeIa are not considered), mixed with
hot evaporated ISM, are located inside a large cavity which extends
above (and below) the galactic disc. Typical parameters of this cavity
are the linear size $d>1$ kpc, the density $n_{\rm c}=10^{-2}$
cm$^{-3}$ and the temperature $T_{\rm c}=10^6$ K. After the last SN
explosion, strong radiative losses occur. However, given the density
and temperature fluctuations in the hot medium, cooling acts in a
differential way. This leads to condensation of the metal-rich gas
into small molecular droplets ($R_{\rm c}\sim 0.1$ pc, $M_{\rm c}\sim
1$ $M_{\odot}$) able to fall back and settle on to the
disc of the galaxy. With the next exploding stellar generation, the
droplets are dissociated and disrupted, and their gas is eventually
mixed in the HII regions.

Here we point out some $caveat$ concerning this scenario which
should be taken into account.
\par\noindent
\smallskip

{\it Thermal conduction} - Thermal conduction, if not impeded by
magnetic fields and/or plasma instabilities, introduce, together with
radiative losses, the characteristic Field length
(Begelman \& McKee 1990, Lin \& Murray 2000)

$$
\lambda_{\rm F}=\left ({3\kappa(T)T \over n^2\beta 
\zeta \Lambda(T)}\right )^{1/2}
$$ 
\noindent 
where $\kappa(T)=6\times 10^{-7}T^{2.5}$ erg cm$^{-1}$ K$^{-1}$ is the
classical thermal conductivity (Spitzer 1956) and $n$ the gas
density. The cooling rate scales linearly with the metal content
$\zeta$ of the gas ($\zeta=1$ for solar abundance). The parameter
$\beta$ takes into account the possibility that the cooling gas may be
out of ionization equilibrium; Borkowski, Balbus \& Fristrom (1990)
have shown that $\beta$ may be as high as 10 through conductive
fronts.
 
Clouds undergo evaporation unmodified by radiative losses if they are
sufficiently small, $R_{\rm c}<\lambda_{\rm F}$. Assuming $T_{\rm
c}=10^6$ K and $n_{\rm c} =10^{-2}$ cm$^{-3}$, Tenorio-Tagle obtains
$R_{\rm c}< 10$ pc for the radius of the overdense zones at the
beginning of their implosion.  Adopting $\beta=\zeta=1$ and
$\Lambda=1.6\times 10^{-18}T^{-0.7}$ ergs cm$^3$ s$^{-1}$ for $10^5$ K
$\leq T \leq$ 10$^{7.5}$ K (Mac Low \& McCray 1988), we obtain
$\lambda_{\rm F}\sim$ 138 pc. Even assuming $\beta=10$, $\lambda_{\rm
F}\sim 44$ pc remains larger than $R_{\rm c}$. Thus the rate of
conductive heat input exceeds that of the radiative losses and the cloud
collapse is inhibited. This result derives from the general property
of the evaporation to stabilize thermal instabilities (Begelman \&
McKee 1990). Actually, Tenorio-Tagle obtained $R_{\rm c}\sim 10$ pc as
a $lower$ limit for the radius value of clouds able to implode. As a
consequence, perturbations with a size larger than $\lambda_{\rm F}$
can actually grow. A non negligible fraction of gas mass would
condense only for a rather flat size spectrum of the fluctuations.

We point out that, even if droplets actually form, they evaporate in a
time $\tau_{\rm e}=M_{\rm c}/\dot M$. Droplets have rather small final
radii of order 0.1 pc (Tenorio-Tagle 1996). In this case they undergo
the satured evaporation $\dot M=1.22\times 10^{-14}T^{2.5}R_{\rm
c}\sigma^{-5/8}$ g s$^{-1}$, where $\sigma=3\times
10^{18}(T/1.54\times 10^7)^2/(nR_{\rm c})$ is the saturation parameter
(Cowie, McKee \& Ostriker 1981). For $M_{\rm c}\sim 1$ $M_{\odot}$ we
have $\tau_{\rm e}\sim 40$ Myr, comparable to the dynamical time
$\tau_{\rm d}$ (see below).  Thus all the droplet material, or at
least a large fraction of it, returns into the hot phase before
reaching the galactic plane.  \par\noindent
\smallskip

{\it Drag disruption} - Suppose that thermal conduction is impeded and
that droplets with final $R_{\rm c}\sim 0.1$ pc actually form and fall
toward the galactic plane. During their descent the droplets
experience a drag reaching a terminal speed $V_{\rm t} \sim (\chi
R_{\rm c} /d)^{0.5}V_{\rm c}$, where $\chi=10^5$ is the ratio of the
droplet density to the hot gas density, $d\sim 1$ kpc is the droplet
distance to the galactic plane, and $V_{\rm c}$ is the circular
velocity of the halo potential. The motion of the droplet relative to
the hot gas leads to mass loss through Kelvin-Helmholtz
instability. For wavelengths $\lambda \sim R_{\rm c}$ the stripping
time-scale is $\tau_{\rm s}\sim R_{\rm c} \chi^{0.5}/V_{\rm
t}=\tau_{\rm d}(R_{\rm c}/d)^{0.5}$ (cf. Lin \& Murray 2000), where
$\tau_{\rm d}=d/V_{\rm c}$ is the dynamical time. The droplets are
thus disintegrated before they settle to the galactic plane, returning
to the hot diluted phase. Larger droplets may not be able to attain
their terminal velocity, but even in this case we have $\tau_{\rm
s}/\tau_{\rm d}\sim \chi ^{0.5}R_{\rm c}/d<1$.

As an aside, we note that, if
a large fraction of the hot gas becomes locked into droplets, the
pressure of the remaining diluted phase reduces and the cavity
shrinks.  Whether the bubble deflates slowly or suddenly and producing
turbulence depends on the droplet formation efficiency. Thus,
the scenario depicted by Tenorio-Tagle of a nearly steady hot cavity
with size of $> 1$ kpc waiting for the onset of radiative cooling
could be incorrect. Part of the droplets could be overrun by the edge of
the imploding bubble, undergoing an even faster stripping.

\par\noindent 
\smallskip
%

Let us consider now the results shown in the present paper. Contrary
to the scenario sketched above, the ejecta cool rapidly without
leaving the galaxy (until the break-out, which occurs at late times)
and without undertaking a long journey before mixing with the ISM. How
much these results are reliable? Mac Low \& McCray (1988) showed that
a conductive bubble expanding in an uniform medium becomes radiative
(i.e. radiates an energy comparable to the thermal energy content of
the shocked wind) after a time:
$$t_{\rm R}\sim 16(\beta \zeta)^{-35/22}L_{38}^{3/11}n^{-8/11}\; {\rm Myr,}$$
\noindent
when the cavity radius is:
$$R_{\rm R}\sim 350(\beta \zeta)^{-27/22}L_{38}^{4/11}n^{-7/11}\; {\rm
pc.}$$ 
\noindent
For $t>t_{\rm R}$ the bubble goes out of the energy conserving
regime, although a fully momentum conserving regime is never
attained. Considering model M1, we assume $\beta=\zeta=1$, $L_{38}=2$
and $n=1.8$ cm$^{-3}$, and we obtain $t_{\rm R}=12.6$ Myr and $R_{\rm
R}=310$ pc. Thus, in the case of an uniform unperturbed medium the
bubble interior would cool quite early, when it is still well inside
the galaxy.

Mac Low \& McCray also considered the expansion in a stratified medium.
Equating the radius of a spherical bubble to approximatively one scale height
$H$ they define the dynamical time
$$t_{\rm D}\sim H^{5/3}(\rho/L_{\rm w})^{1/3}.$$
\noindent
Then the ratio of cooling to dynamical time-scales is:
\begin{equation}
{t_{\rm R}\over t_{\rm D}}=8.22n^{-35/33}L_{38}^{20/33}
\left ({H\over 100 {\rm pc}}\right )^{-5/3}(\beta \zeta)^{-35/22},
\end{equation}
\noindent
(note that the numerical coefficient in this expression slightly
differs from that obtained by Mac Low \& McCray).  For model M1 we
obtain $t_{\rm R}/ t_{\rm D}\sim 1.08$ during the SNII stage.  Thus a
non negligible fraction of the wind luminosity is radiated away (see
also Fig. 9), and the break out does not occur. For M3, where $n\sim
0.5$, we have $t_{\rm R}/ t_{\rm D}\sim 4.2$, and the situation, in
principle, is less clear-cut (see, however, below in this section).

Although models M1, M2, M3 do not take explicitly into account heat
conduction, yet they obtain results according to the above
scenario. In fact, as pointed out in section 3.1.1, numerical
diffusion simulates thermal conduction originating spurious radiative
losses which otherwise would be absent. Of course, this spurious
cooling does not reproduce $quantitatively$ the same amount of
radiation lost through a real heat conduction front, and the fraction
of cold ejecta obtained in our models could be larger than the correct
one. Some algorithms may be conveniently adopted to reduce this
effect. Consistent advection (Stone \& Norman 1992) is implemented in
our code and helps in reducing somewhat the diffusion, making it
consistent for the all advecteded quantities (mass, momentum,
energy). We also made tests modifing the cooling algorithm in presence
of unresolved contacts, following Stone \& Norman (1993). Although the
fraction of cold ejecta reduces of 15 per cent, most of the metals ($\sim$
80 per cent) remains cold. We point out, however, that in presence of an
unresolved conduction front, the above algorithm may lead to an
excessive reduction of the radiative losses (see below).

In any case the intrinsic diffusion of the code may be alleviated but
not eliminated by algorithms as those described. In principle, more
realistic models can be done explicitly adding the physical terms
which produce diffusion.  For this reason we also ran model MC, where
the heat transfer is included. In this model the amount of cold metals
does not change appreciably. However, although model MC is useful to
understand the stabilizing effect of heat conduction on a turbulent
flow, it turns out to be inadequate to obtain the correct cooling rate
at the conduction front. Consider the temperature profile of a
``standard" bubble $T_{\rm b}(1-r/r_{\rm b})^{2/5}$ (Weaver et al. 
1977), where $T_{\rm b}=10^6$ K is the central temperature, and
$r_{\rm b}=300$ pc is the bubble radius (cf Fig. 7). The cooling curve
maximum occurs at $T\sim 2\times 10^5$ K. This temperature is found at
$r/r_{\rm b}=0.98$, i.e. at a distance of 6 pc to the cold shell. At a
distance of 300 pc the mesh size is $\sim 15$ pc, and the conduction
front is not resolved properly. Thus, we also ran models MCH and MCHH
(not shown here) with heat conduction and with an uniform grid with
mesh size of 2 and 1 pc respectively. These models were computed only
up to the end of the SNII activity because of the large number of grid
points involved.

The four panels in Fig. 10 show the profiles along the galactic plane
($z=0$) of several quantities for models M1, MC, MCH and MCHH at
nearly 30 Myr.  As expected, the resolution of the temperature profile
is not improved in MC and the temperature jump remains unresolved. In
MCH, instead, this jump extends over 2-3 meshes and in MCHH over 4-5,
as expected. The fraction of cold ejecta is 0.95, 0.95, 0.93, 0.92 for
M1, MC, MCH, MCHH respectively.  Although the greater accuracy, the
fraction of cold ejecta in MCH and MCHH is only few percent lower than
in M1. This occurs because the bubble is ``genuinely'' radiative, and
an high spatial resolution may retard a little the cooling, but cannot
avoid it.

We stress once more (cf. section 3.1.1) that it is very difficult to
give the correct description of the contact discontinuity at the outer
edge of the hot cavity, for the presence of complex hydrodynamical
phenomena occurring there. These phenomena tend to produce a finite
thickness of the contact, giving rise to a substantial cooling
otherwise absent. The correct evaluation of such a cooling is however
very difficult to assess. Even restricting ourselves on the simple
case of heat conduction, the possible lack of ionizing equilibrium and
the numerical diffusion both increase radiative losses. The first
effect is physical and could be described solving step by step the
time dependent set of ionization equation. The second is a spurious
result due to numerical diffusion of the code which must be reduced as
much as possible. We believe that our convergence test (cf. fig. 10),
as well other 1D tests not shown here, indicates that our results are
significative. We are aware of the fact that more refined simulations
may produce somewhat different values of the cooled mass of the
ejecta, but we think that the large fraction of resulting cooled
ejecta is a genuine result.

\begin{figure*}
\centering
\vspace{-0.2cm}
\epsfig{file=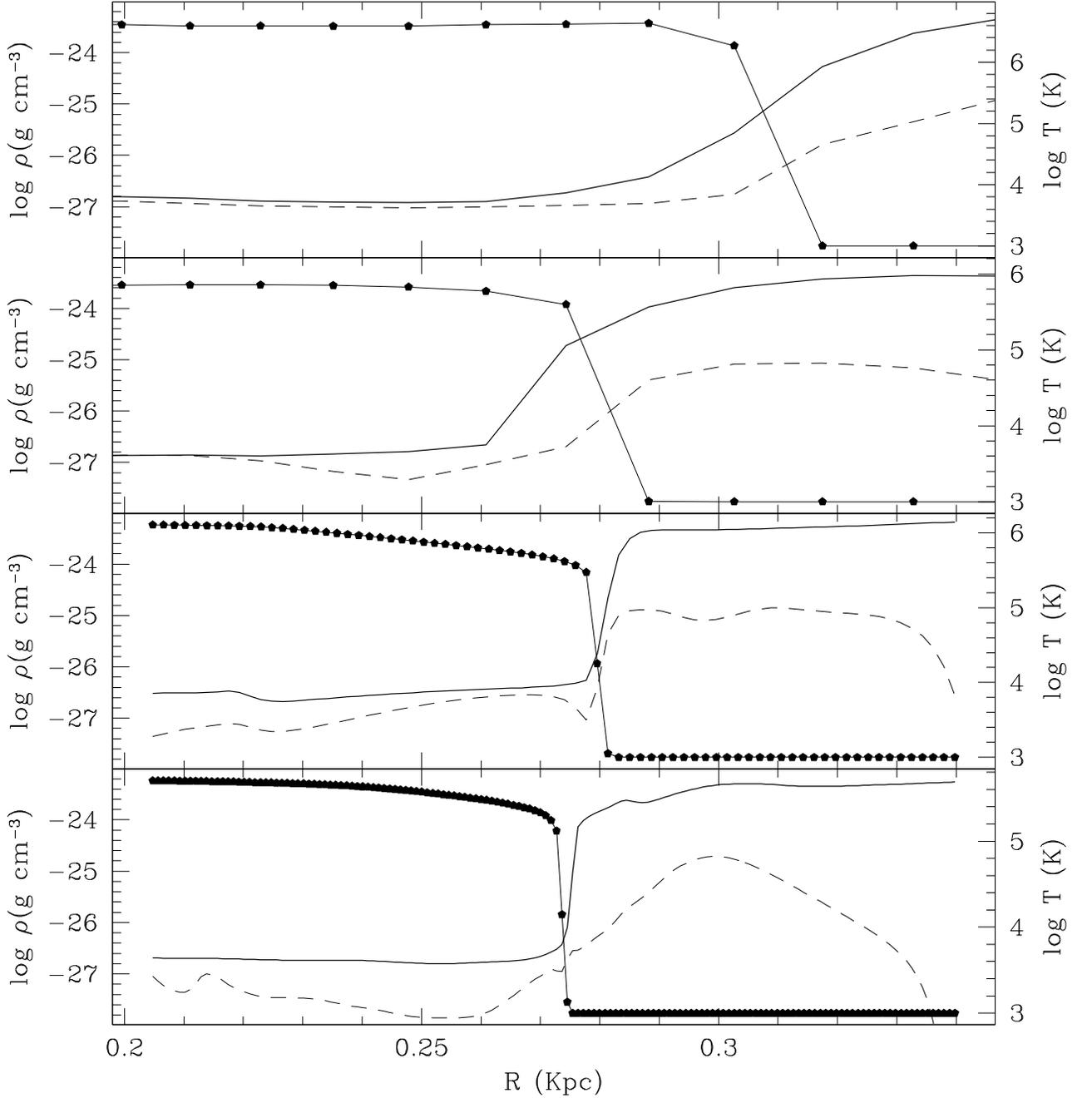,height=18.5cm,width=18.5cm}
\caption[]{\label{fig:fig 10} Density (heavy solid lines), ejecta
(dashed lines) and temperature (light solid lines) profiles for models
M1 (first panel starting from top), MC (second panel), MCH (third
panel) and MCHH (bottom panel) near the conduction front, along the
$R$ direction, after $\sim$ 30 Myr.  The diamonds superimposed to the
temperature profiles, indicate the mesh points of the grid for each
numerical simulation. }
\end{figure*}

As pointed out above, equation (21) gives an ambiguous prevision about
the behaviour of M3.  We therefore ran also this model on a uniform
grid with 2 pc resolution (up to $t=30$ Myr) adding heat
conduction. The overall dynamics of the superbubble remains the same,
and the fraction of cold ejecta turns out to be 6 per cent lower than
in the low resolution model.  We thus conclude that the results
obtained by our models are reliable.

At the end of this section we mention the effects expected in the case
of a non-uniform initial ISM. Actually, gas clouds are embedded in the
pervasive diffuse gas of the galaxy.  However, a correct numerical
treatment of the interaction between these clouds and the ambient gas
introduces enormous complications in the description of the involved
physics and needs full 3D computations with an huge number of
meshes. We here just make some simple considerations following McKee,
Van Buren \& Lazareff (1984).  These authors describe the behaviour of
a bubble generated by an O star and expanding in a cloudy
medium. Because of the flux of ionizing photons emitted by the star,
the nearby clouds undergo photoevaporation and accelerate away through
the rocket effect. Essentially no cloud survives up to a radius
$R_{\rm h}$, and the gas density inside this radius increases to a
value 0.5$n_{\rm m}$, i.e. half of the mean density the cloud gas
would have if it were homogenized. The wind bubble evolution depends
on the value of $L_*=L_{\rm w}/L_{\rm St}$, where $L_{\rm
St}=1.26\times 10^{36}(S_{49}^2/n_{\rm m})^{1/3}$ and $S_{49}$ is the
rate at which the star emits ionizing photons in units of 10$^{49}$
s$^{-1}$. For weak winds ($L_*\ll 1$) the bubble radius is smaller
than $R_{\rm h}$ and it evolves ``normally'' (Weaver et al.
1977). For moderate winds ($L_*\sim 1$) the bubble expands to the edge
of the cloud distribution because photo-evaporated gas induces
radiative losses reducing the pressure. Finally, for $L_*\gg 1$ the
bubble rapidly engulfs a number of clouds and radiates away most of
its internal energy. This scenario cannot be directly applied to a
star burst as a whole. In fact clouds are also present inside the star
formation region, partially screening the flux of ionizing photons
escaping from this region.  However, even assuming that this effect is
negligible, in our model $L_*\sim 3$. We in fact computed $S_{\rm 49}$
running remotly at the site www.stsci.edu/science/starburst99
(Leitherer et al. 1999) a burst model tailored on that assumed here. 
The number of UV photons produced by massive stars remains nearly constant
($S_{49}=584$ s$^{-1}$) up to 5 Myr, and then drops as $t^{-4}$. For
$n_{\rm m}=1.8$ cm$^{-3}$ close to the galactic disc, we have $L_{\rm
St}=7\times 10^{37}$ erg s$^{-1}$, lower than $L_{\rm w}$. Of
course, along the $z$ direction $R_{\rm h}$ could move much further,
but the ionizing flux starts to decline rather soon, and the cloud
distribution remains nearly unaffected. Thus the bubble is expected to
become radiative sooner than in the case of a smoothly distributed ISM.

In conclusion, in the scenario of our models, most of the metals
actually cools off in a few Myr.  It is worth noting that this result
depends essentially on the assumption of a low heating efficiency of
SNIIs. In the model similar to M1, but with $\eta_{\rm II}=1$
(sect. 2.3.2), the break out quickly occurs and the metals have no
time to cool. In fact, following Eq. (21), in this case the bubble results to
be adiabatic and not radiative. Thus the SN efficiency value is
crucial, and a future paper will be devoted to it (D'Ercole \&
Melioli, in preparation).

\subsection{Chemical results}

In Fig. 10 is shown the evolution of the element abundances for the
models M1B, M2B and M3B. In this figure, the evolution of the masses
and the abundances in the form of the various elements is shown.  It
is interesting to note that the mass of the lost metals for the model
M3B is larger than that retained from the galaxy, whereas it is the
contrary for models M1B and M2B. This means that the initial
conditions, namely the assumed burst luminosity and gas mass in the
galaxy, are playing a very crucial role in the development and
evolution of the galactic wind. The abundances are calculated as [Z/H]
, where $Z$ indicates the abundance of the following elements, O, C, N,
Mg, Si and Fe, relative to the solar abundances of Anders \& Grevesse
(1989), and as 12+log(Z/H), which is the notation normally used for
the abundances in extragalactic H\,{\sc ii} regions. These abundances are
derived in the following way: for the galactic abundances we have
averaged in the previously defined galactic region (approximatively an
ellipsoid with major semi-axis of 1 Kpc and minor semi-axis of 730
pc), whereas for the abundances leaving the galaxy the integral is
made over the rest of the grid.

The $\alpha$-elements show a very similar evolution and this is due
mostly to their common origin. The $\alpha$-elements are, in fact,
mainly produced by massive stars (see Figg. 2--4), and thus their
abundances inside the galaxy grow in the first 6-7 Myr -- the time
interval where most of massive stars die -- then show a slowly
decreasing trend, with however a little maximum around 100 Myr.  This
behaviour is related to the dynamics of gas flows described before.

The evolution of iron is not too different with respect to the
evolution of the $\alpha$-elements although this element is
substantially produced by SNeIa. This is due to the fact that at the
end of our simulations the iron produced by SNIa is only $\sim$
30 per cent of the total, because the bulk of SNeIa appears at later
times.  The evolution of N shows a sharp increase at around 29 Myr
which corresponds to the lifetime of a 8 M$_{ \odot}$ star, which is
the first star producing a substantial amount of N (of primary
origin). For times less than 29 Myr the N production is negligible.

\begin{figure*}
\centering
\epsfig{file=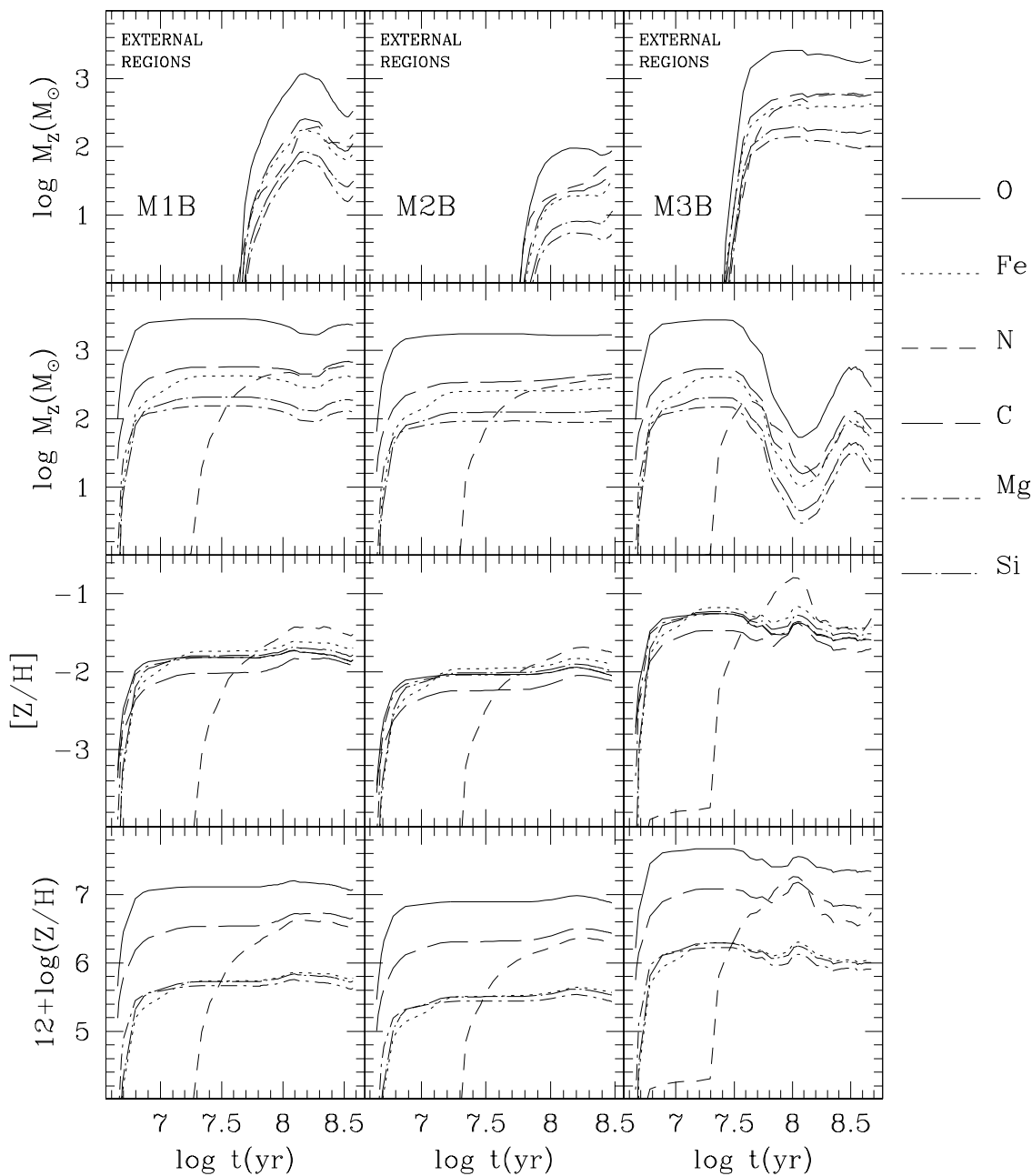,height=18cm, width=18cm}
\caption[]{ Time evolution of several quantities for the models M1B,
M2B and M3B (first, second and third column, respectively). 
The upper pannels show the evolution of the mass of
various elements outside the galaxy, while the masses inside the galaxy are 
shown by the second raw of pannels. The third and the last raws of panels 
illustrate the behaviour of the ISM abundances (relative to the sun and 
by number, respectively).} 
\end{figure*} 

\section{Discussion}

We want to compare now our models with observational data 
found in literature for the galaxy IZw18.
\subsection{Morphology and dynamics}

\par
\hbox{}
\begin{table*}
\begin{flushleft}
\caption[]{Predicted abundances in the galactic region after 31 Myr
for models M1, M2, M3. We also show the results for model MC (with 
nucleosynthetic prescriptions B), in order to emphasize the
similarities with model M1B. Only models M1 and MC should be compared with
IZw18. These values are compared with some abundances found in
literature for IZw18.}
\begin{tabular}{cccc|c}
\noalign{\smallskip}
\hline
\noalign{\smallskip}
12+log(O/H) \hspace{0.6cm}& 
log C/O   & 
log N/O   & 
\hspace{0.5cm} log Si/O \hspace{0.5cm}  & 
Fonts \\
\noalign{\smallskip}
\hline\noalign{\smallskip}
$7.12$ & $-0.47$ & $-5.24$ & $-1.40$ & M1A\\
\underline{$7.12$} & \underline{$-0.58$} & \underline{$-1.52$} & 
\underline{$-1.40$} & \underline{M1B}\\
$7.51$ & $-0.57$ & $-1.46$ & $-1.29$ & M1C\\
$7.51$ & $-0.63$ & $-1.33$ & $-1.29$ & M1D\\
$6.90$ & $-0.49$ & $-5.02$ & $-1.40$ & M2A\\
$6.90$ & $-0.58$ & $-1.61$ & $-1.40$ & M2B\\ 
$7.36$ & $-0.54$ & $-1.31$ & $-1.30$ & M2C\\
$7.36$ & $-0.58$ & $-1.24$ & $-1.30$ & M2D\\
$7.67$ & $-0.49$ & $-5.22$ & $-1.38$ & M3A\\
$7.67$ & $-0.59$ & $-1.60$ & $-1.38$ & M3B\\
$7.97$ & $-0.64$ & $-1.90$ & $-1.26$ & M3C\\
$7.97$ & $-0.71$ & $-1.61$ & $-1.26$ & M3D\\
$7.12$ & $-0.58$ & $-1.49$ & $-1.39$ & MC\\
\noalign{\bigskip}
$7.24$ & $-0.54$ & $-1.54$ & - & DH\\
$7.17/7.26$ & - & $-1.54/-1.60$ & - & SK\\
$7.17/7.26$ & $-0.63/-0.56$ & $-1.56/-1.60$ & - & G97\\
- & - & - & $-1.52$ & G95\\
\noalign{\smallskip}
\hline
\end{tabular}

References: DH: Dufour \& Hester (1990); SK: Skillman \& Kennicut (1993); 
G97: Garnett et al. (1997); G95: Garnett et al. (1995).
\end{flushleft}
\end{table*}  

To compare our dynamical results we should first summarize the
structural and dynamical properties of IZw18.  IZw18 has a
`peanut-shaped' main body, consisting of two starbursting regions
(Dufour et al. 1996). There are also two H\,{\sc ii} regions (also
called NW and SE), associated with the main body, but shifted $\sim$ 1
arcsec east of the brightest continuum emission (Martin 1996).  The
H$\alpha$ emission is bipolar-shaped along a direction orthogonal to
the main body, and show clear evidences of shell structures. In fact a
prominent shell stretches 15 arcsec (720 pc) north-northeast from the
northwest H\,{\sc ii} region and bright H$\alpha$ emission extends
symmetrically south-southwest from the NW region (Martin
1996). Moreover a partial shell of 3.6 arcsec of diameter (173 pc)
protrudes from the north-west side. The H\,{\sc i} velocity field (Van
Zee et al. 1998; Viallefond et al. 1987) shows a significant velocity
gradient along minor axis, suggesting a flow in this direction (Meurer
1991).

We first note that the distance of 720 pc between the shells and the
NW H\,{\sc ii} region is quite compatible with models M1, M2, M3 (in
these models, after 31 Myr, the bubble has covered a distance of
$\sim$ 600--800 pc along the $z$-axis). Martin, starting from
geometrical considerations, found a shell speed of 35--60 Km s$^{-1}$,
and in our model M1, after 31 Myr, the velocity of the outer shock
along the $z$-axis is approximatively 30 Km s$^{-1}$, in agreement
with the observations.  We note instead that the model MC does not
develop an outflow along the $z$ direction (see Fig. 8a), and this is
due to the fact that the bubble never extends beyond $H_{\rm eff}$, as
explained in section 3.1.4. Actually, the uncertainties on the real
value of $H_{\rm eff}$ in IZw18 are significant and a small reduction
of $H_{\rm eff}$ could lead to the formation of an outflow also for
model MC.

These comparisons between observation and theory depend strongly on
the adopted distance of IZw18. In a recent work, Izotov et al. (1999)
found that the distance of IZw18 should be at least 20 Mpc, twice the
distance generally adopted for this galaxy.  With this new distance,
Color-Magnitude Diagram (CMD) studies give an age, derived from the
main sequence turn-off, of 5 Myr for the main body. They suppose that
the star formation in the main body has started $\sim$ 20 Myr ago in
the NW edge, propagating then toward the SE direction and then
triggering the main starburst $\sim$ 5 Myr ago. This estimate is also
consistent with the stellar population analysis of Hunter \& Thronson
(1995).

Here we have assumed a coeval stellar population, so we cannot
correctly verify the hypothesis of Izotov et al., but in our models a
slightly pre-enriched burst with an age of 5 Myr cannot account for
the observed abundances and the nearly flat metallicity gradient
observed in IZw18 (see next section).  Moreover, with the new
estimated distance the above-mentioned shell structures double their
dimensions and it is hard to reproduce these morphological features
with a burst 5 Myr old.

\subsection{Chemical abundances}

In Table 4 the abundance ratios log(C/O), log(N/O), log(Si/O) as well
as 12+log(O/H) predicted by our models are reported. At the beginning
of section 3 we described several factors which affect the metal
content of the ISM.  From an inspection of Fig. 11 and Table 4 we
conclude that a reduction in the burst luminosity produces a reduction
in the total abundances, while a decrease in the ISM mass leads to an
increase of the metallicity.

In Table 4 the observed abundances of IZw18 are also
reported. However, only the model M1 (cases A,B,C,D) should be
compared with IZw18, since the total mass and the gas mass of this
model have been chosen to match this galaxy. Before to continue the
discussion on our results, we point out that the abundances shown in
Table 4 refer to the whole gas into the galaxy, while the comparison
with the data should be valid only for the cold ($T<2\times 10^4$ K)
phase.  In fact, chemical composition of stellar winds and supernovae
ejecta are mainly measured through the relative intensities of visual
[O\,{\sc ii}], [O\,{\sc iii}], [S\,{\sc ii}] and [N\,{\sc ii}]
forbidden lines compared to H and He recombination lines, but this
approach is sensitive only to emission from warm, photo- or
shock-ionized gas at $\sim$ 10000 K (Kobulnicky \& Skillman 1997).

However, although the metal abundances in the hot regions are quite
large, reaching also extrasolar values (cf. Fig. 14), in our models
the majority of the metals is in the cold gas phase.  Thus the
abundances in Table 4 are essentially the same as those in the cold
gas, and their comparison with the observed metallicities is
meaningfull. Note that, in making such a comparison, we suggest that
the present time burst in IZw18 can be responsible of the observed
chemical enrichment in the H\,{\sc ii} regions. This is in agreement
with the arguments illustrated in section 3.2, and at variance with
previous suggestions of different authors (see e.g. Larsen et
al. 2000).  Recent observations (Pettini \& Lipman 1995; Van Zee et
al. 1998), although uncertain, indicate an oxygen abundance in the
H\,{\sc i} regions of IZw18 comparable with that in the H\,{\sc ii}
regions, in agreement with our predictions.

Table 4 reports the abundances of our models after 31 Myr. The
lifetime of a 8 M$_{\odot}$ star is approximatively 29 Myr, according
to eq. (14). Therefore, since at $Z=0$ secondary N is not produced and
primary N from massive stars is negligible, only for ages larger than
29 Myr we can expect some N which is the one produced in a primary
fashion by IMS during the third dredge-up episode.  As one can see
from Table 4, the abundances and abundance ratios predicted by the
yields for $Z=0$ and $\alpha_{RV}=1.5$ (model M1B) are in good
agreement with those measured in IZw18, thus we could conclude that
the abundances in this galaxy are compatible with only one burst, the
first, but only if the burst age is of the order of 31 Myr. In fact,
for times shorter than that the N abundance is too low and for times
larger the agreement worsens.

The abundance of C and particularly the predicted C/O ratio in model
M1B is in very good agreement with observations at variance with
previous works (Kunth et al. 1995).  The difference between the low
abundance of $^{12}$C predicted by Kunth et al. (1995) and here is, in
our opinion, due to the fact that the total amount of stars produced
there was smaller ($M_{\rm burst} = 2 \times 10^{5}$ M$_{\odot}$) and
less in agreement with the observations than that produced here
($M_{\rm burst} = 6\times 10^{5}$ M$_{\odot}$), therefore we predict
higher abundances for all the elements. In addition, the yields for
massive stars used here are different from those used in Kunth et
al. (1994) (those of Woosley 1987).  It is also worth mentioning that
the estimated ages for the present burst in IZw18 are between 15 and
27 Myr (Martin 1996) in good agreement with our suggestion, although
other authors suggest ages as short as 5 Myr (Izotov et al. 1999;
Stasi\`nska \& Schaerer 1999).

In order to see if we can exclude a previous burst besides the present
one in IZw18, as suggested by previous papers (Aloisi et al. 1999;
Kunth et al. 1995), or a recent burst coupled with a low but
continuous star formation (Legrand 1999), we computed the expected ISM
abundances for a preenriched gas with $Z=0.01$ Z$_{\odot}$ and we show
in Table 4 the abundances for this case at an age of 31 Myr (cases C
and D). The results for the C case show that at an age of 31 Myr the
abundance of oxygen is too high; the results at the end of the
simulation (375 Myr for model M1) give better values for oxygen and
N/O but they predict a too high C/O. If one assumes then $Z=0.01$
Z$_{\odot}$ and $\alpha_{RV}=1.5$ (case D), the agreement worsens at
any age since one predicts a too high N/O ratio while the rest is
practically unchanged. Therefore, we have two considerations: first,
the single first burst hypothesis seems to give the best agreement
with observations as long as primary N production in IMS is
considered; second, we cannot really exclude a previous burst before
the present one, or at least we cannot exclude a burst which enriched
only sligthly the ISM. In other words, the preenrichment should be
less than $Z=0.01$ Z$_{\odot}$. From the previous discussion, it
arises that the best age for the burst in IZw18 should be around 31
Myr.

\begin{figure}
\centering
\epsfig{file=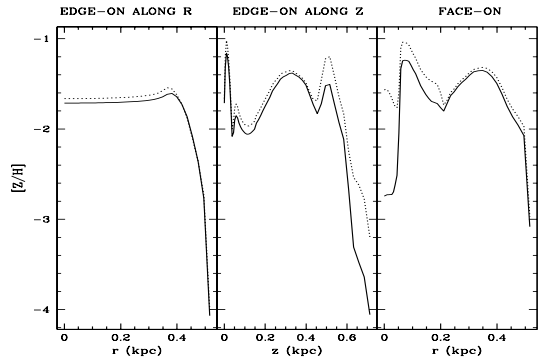,height=5.5cm,width=7.5cm}
\caption[]{\label{fig:fig 12} Abundance gradients for oxygen 
(solid line) and iron (dotted line) after 31 Myr, for the model M1B.}
\end{figure}     

At this age the abundance gradient in a region of 600 pc is almost
flat, at least if the galaxy is seen edge-on (see Fig. 12), in
agreement with what is observed in IZw18 (Legrand 1999). Actually, if
a bipolar-shaped expanding bubble is present in IZw18, the inclination
of the symmetry axis with respect to the normal to the observer would
be very small (Martin 1996, suggested an inclination of
10$^{\circ}$). 

\begin{figure*}
\centering
\epsfig{file=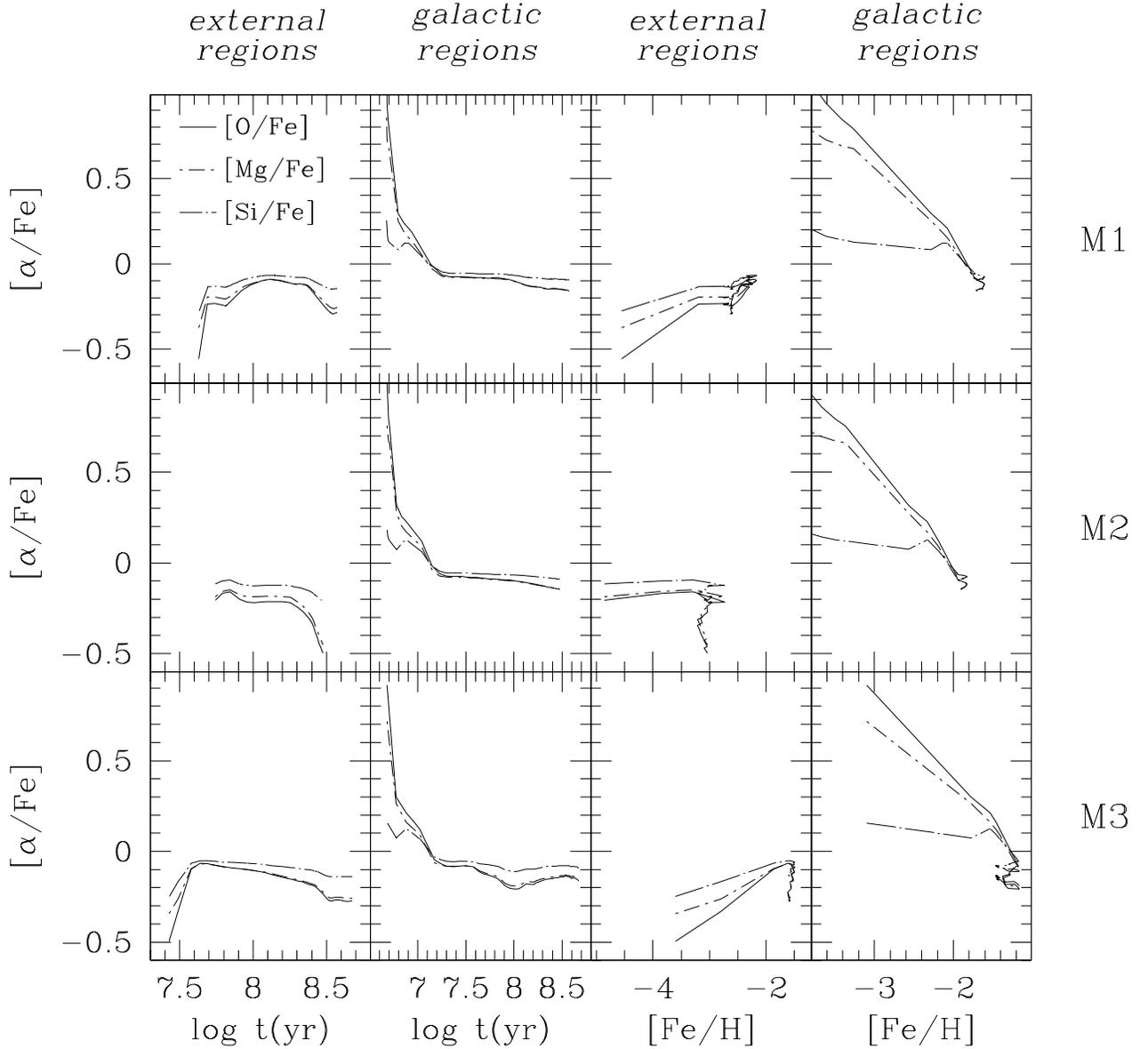,height=17.7cm,width=18cm}
\caption[]{\label{fig:fig 13} Predicted [$\alpha$/Fe] vs. time and vs. [Fe/H] 
for models M1B, M2B, M3B (first, second and third raw respectively) for both 
expelled gas and ISM}
\end{figure*}

Finally, in Fig. 13 we show the predicted [$\alpha$/Fe] vs. time and
vs. [Fe/H] for the gas inside and outside the galaxy, corresponding to
the results of Models M1B, M2B and M3B.  The interesting feature of
this figure is that the [$\alpha$/Fe] ratios in the gas outside the
galaxy are lower than those in the gas inside the galaxy. This is due
to the fact that Fe, in particular that produced by type Ia SNe, is
lost more efficiently than $\alpha$ elements.

\begin{figure*}
\centering
\epsfig{file=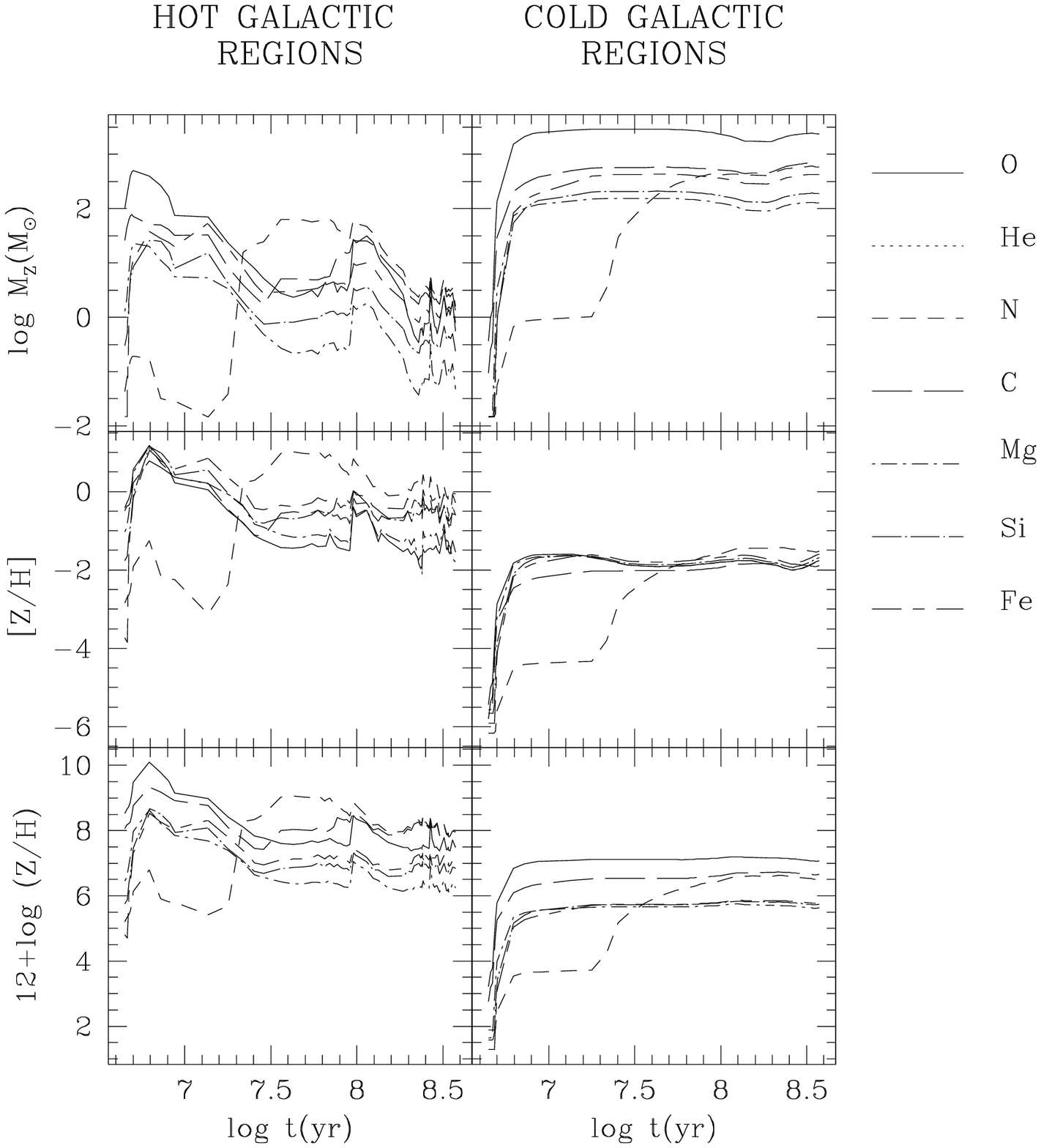,height=15.5cm,width=15.5cm}
\caption[]{\label{fig:fig 14} Abundance evolution in hot and cold 
regions for the model M1B. The threshold temperature is $2\times 
10^4$ K.}
\end{figure*}

We shall also note in these two figures the peculiar behaviour of
silicon relative to the other $\alpha$-elements. Silicon is, in fact,
synthetized by SNeIa at variance with what happens for O and Mg, which
are mainly produced by SNeII (see Gibson et al. 1997).  This is the
reason for the relatively low and flat [Si/Fe] ratio as a function of
[Fe/H].

\section{Conclusions}

We have studied the dynamical and chemical evolution of a dwarf galaxy
as due to the effect of a single, instantaneous, point-like starburst
occurring in its centre. We adopted galactic structural parameters
which resemble those of IZw18, the most unevolved dwarf blue compact
galaxy known locally.  We ran different models, which differ
for the burst luminosity and the ISM mass. We considered the mass and
energy inputs from the single low and IMS, the SNeII and the SNeIa
(white dwarfs in binary systems) and we followed the evolution of the
gas and its chemical abundances (H, He, C, N, O, Mg, Si and Fe) in
space and time for several hundreds Myr from the burst.

Our results can be summarized as follows:

\begin{enumerate}

\item The starburst can inject enough energy into the ISM to trigger
a metal enriched galactic wind: the metals synthetized and ejected
through supernova explosions leave the galaxy more easily than the
unprocessed gas.  This result is not new since it has already been
suggested by previous works (e. g. MF and DB).  However, our new
result is that the SNIa products have the largest ejection efficiency
(more than the products of type II SNe), with the consequence that the
[$\alpha$/Fe] ratios in the gas outside the galaxy are predicted to be
lower than those inside. This is due to the fact that SNeIa produce a
substantial fraction of iron. 

\item The energy injection in the ISM by SNII has a rather low efficiency.
Instead, the energetic contribution of SNeIa, in spite of their
relatively small number (a total of 240 SNeIa against 4800 SNeII), has
important consequences in the dynamical behaviour of the galaxy. Since
the SNIa explosions occur in a medium heated and diluted by the
previous activity of SNeII, the thermal energy of the explosions is
easily converted into kinetic energy and so the gas reaches quickly
the outer regions of the bubble along the galactic chimney. DB showed
that after the end of SNII activity, a fraction of the gas tends to
recollapse toward the central region of the galaxy, achieving the
threshold density for a new star formation event ($N({\rm H}\,{\sc i})\simgt
10^{21}\;{\rm cm}^{-2}$, Skillman et al. 1988; Sait\=o et al. 1992)
in $\sim$ 0.5--1 Gyr. With the energetic contribution of SNeIa, it is
not possible to reach, at least for the time considered in our
simulation, such a threshold in column density (it can be obtained
only after many Gyr).

\item One single burst, occurring in a primordial gas, with an age of
$\sim$ 31 Myr reproduces quite well both the dynamical structures and
the abundances in IZw18. This value is consistent with other
independent age estimates for the burst (Martin 1996).  From the
nucleosynthetic point of view the age of 31 Myr ensures that there is
enough time for the primary N from IMS to be produced and ejected.
The adopted yields for massive stars (Woosley \& Weaver 1995) include
some primary N, but its amount is negligible. This result suggests that
IZw18 is probably experiencing its first major burst of star
formation, although we cannot exclude a previous burst (see Aloisi et
al. 1999) of moderate intensity, which enriched the gas to a
metallicity $Z < 0.01$ Z$_{\odot}$.

\item At variance with previous studies we find that the majority of
metals (in mass) are found in the cold gas. In fact, mainly because of the
low SNII efficiency, the wind superbubble remains several hundreds of
Myr inside the galaxy before the break out occurs. Moreover, the
superbubble becomes radiative after a few Myr, and most of the SNII
ejecta cool without leaving the galaxy (SNIa ejecta, instead, are
shown to be vented more easily). Given the relatively short mixing time, the
abundances predicted by our models for the cold gas are those that
should be compared with the abundances observed in IZw18.  Actually,
they are in very good agreement with the observed ones.  This result
supports the common assumption made in chemical evolution models of an
instantaneous mixing.

\end{enumerate}

Future improvements of this work will include models with continuous
bursts, sequential bursts, as well as a more detailed study
of the formation of the H\,{\sc ii} regions (Recchi et al. in
preparation). 

\section*{Acknowledgements}
We are grateful to Guillermo Tenorio-Tagle and Andrea Ferrara for
useful suggestions and discussions. We also thank the referee whose
suggestions improved the paper. We acknowledge financial support from
the Italian Ministry for University and for Scientific and
Technological Research (MURST).

{}
\end{document}